# Ripple formation and its effect on the multi-scale microstructure of Directed Energy Deposition (DED)-printed 316L components

Himanshu Balhara, Bhaskar Botcha, Sarah Wolff, Satish Bukkapatnam[*]

*Department of Industrial and Systems Engineering, Texas A&M University, College Station, TX 77840, United States*

[*]Corresponding author: satish@tamu.edu

**ABSTRACT:**

An experimental study is presented to characterize the ripple formations in the directed energy deposition (DED) process and to investigate the influence of ripples on the heterogeneous microstructure in the scan direction of DED-printed stainless steel (316L) components. While considerable studies on ripple formation exist in the welding literature, these formations in DED process have not received much attention. Also, little efforts exist on the microstructure along the scan direction (or the scan surface) as compared with the build direction. Experiments consisted of printing 10 mm ×10 mm ×10 mm cubical components on an Optomec LENS® 500 hybrid machine tool under different laser power, scan speed and dwell time combinations, chosen according to a Latin hypercube design. The surface of the scan face of the prints was finished to a Ra < 30 nm and etched with Aqua regia for 90 sec. An optical microscope was employed to observe the microstructure at 4 different scales. The studies suggest a significant influence of ripple formations as well as the ripple width on the solidification front, the dendritic patterns as well as the heterogeneous microstructure.

Keywords: Directed Energy Deposition (DED), Ripples, Microstructure, 316L

## 1. INTRODUCTION

Additive Manufacturing (AM) has rapidly evolved from its inception in the early 1980s for creating visualization prototypes to a $4.1 billion industry with a compounded annual growth rate of 33.8% over 2012-2014 [1], especially for manufacturing custom components. Revenues from AM of metal parts grew by 49.4% to $48.7 million during this period [2]. Powder-fed or a blown powder Directed Energy Deposition (DED) is an AM process employed to fabricate near-net-shape metallic components in a layer-wise deposition mode [3]. It employs a laser head consisting of a focused laser beam coupled with the powder delivery system (see Fig. 1). The laser beam melts the powder to form a molten droplet which is deposited or impinged at the desired location on the substrate [4]. The process holds similarities with plasma arc welding in terms of the use of low effective energy in the range of 100 – 300 J/mm. This leads to low part distortion and reduced mixing and diffusion of materials across layers.

Because of its ability to fabricate freeform components at a much higher material deposition rate compared to other metal AM processes, laser-based DED has been considered for applications in the repair of worn and defective components, such as turbine blades, nuclear reactor parts, and providing hard-facing on different substrate materials [5], [6]. More recent studies show DED process being used for delicate precise component manufacture in the bio-implant and aerospace industry [7].

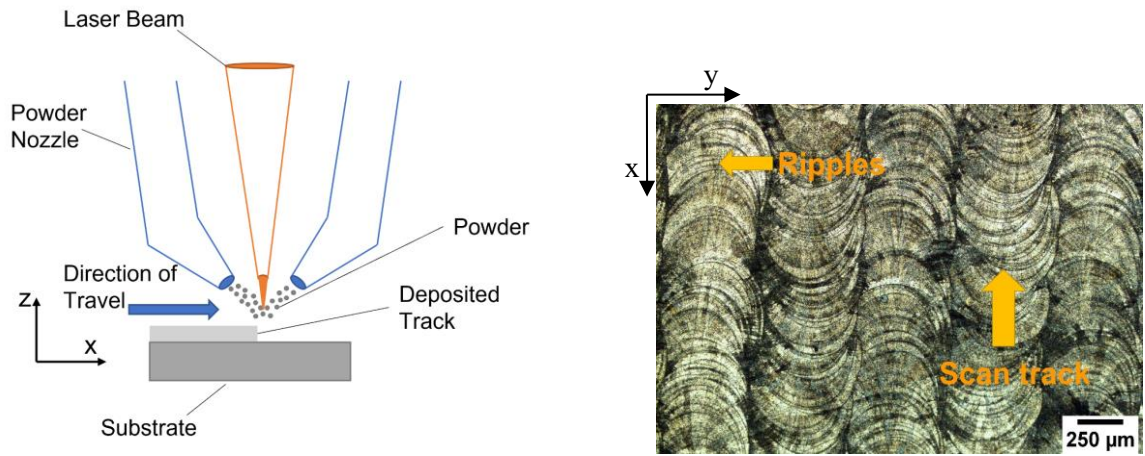

Fig. 1: (a) schematic of a powder fed or blown powder DED process. (b) a representative optical micrograph showing the scan track and ripple formations on DED-printed 316L parts.

The morphology and microstructure of DED-printed components majorly depend on the cooling rate and thermal gradients in the melt pool during the process [8], [9]. Unlike other AM processes, both the thermal and mass-flow fields influence the microstructure, especially in the scan direction. These, in turn, depend on the process parameter settings such as laser power, scan speed, and dwell time [10]. The process offers sharp cooling rates, as high as $10^3 - 10^4$ K/s, which imparts unique microstructures [11]. In DED printed austenitic stainless-steel materials, such as AISI 316L, microstructure patterns are highly heterogeneous and are spread over multiple length scales, extending from a few 100nm up to 1 mm (e.g., [12]–[14]). Additionally, the scan tracks and the ripple structure formed (see Fig. 1b) due to the interaction between the solidification front and the material deposition introduce unique morphological and microstructural patterns spread over these multiple scales. The microstructure patterns, including the grain type, size, orientation, and distribution over multiple scales have a significant, at times, deleterious effect on the mechanical properties and functionality of the DED-printed 316L components. Therefore, characterizing the scan patterns and ripple formations occurring due to the complex thermomechanical interactions under different process conditions becomes quintessential. It helps to delineate the different nonequilibrium phases, their compositions, and spatial distributions to understand the changes in the microstructure within the ripples along the scan direction.

Over 50 studies have reported the various microstructural patterns occurring in the DED-printed 316L components. Avast majority of these studies have focused on capturing the microstructural pattern in the so-called build direction, i.e., along the vertical direction, across multiple layers of material deposition under different process conditions. Only a few studies have discussed the microstructural patterns in scan direction within a single layer (i.e., over a scan plane). The formation of ripples under different process conditions, and how they inform the heterogeneous microstructure in DED has not received much attention in the literature. Understanding these heterogeneous formations within a layer is an important step in augmenting the recent thrusts towards real-time microstructure control in AM processes.

The present experimental study aims to delineate the effect of DED process parameters on the formation of scan tracks and ripples, as well as how they inform the microstructure at different length scales, i.e., from different grain structure formations relative to the ripples and scan patterns. This study also maps the grain structure in the solidification map and identifies different grain clusters at high magnifications. This study is important as understanding the heterogeneity and anisotropy of microstructure over multiple lengths scales ultimately reflects in the properties of DED-printed components. The main contributions of this paper are:

(1) The present optical micrography study focuses on delineating the heterogeneous microstructure on the scan face of DED-printed 316L samples. Very few earlier works studied the microstructure on the scan face. More pertinently, the present work examines the microstructure over four different magnifications to capture the influence of the morphological patterns on the various microstructure clusters.

(2) An experimental study of optical micrographs taken over multiple magnifications capture the effect of process parameters and the resulting thermal and mass-flow history on the geometry of the scan tracks and ripples, and the effect of these morphological formations on the heterogeneous microstructure in the DED-printed 316L. This is perhaps the first study to characterize the ripple formations and their influence on the microstructure in a DED process.

(3) Borrowing from literature on ripple formation in welding processes, we discuss the key physical phenomena that underpin the ripple formation in a DED process. The experimental studies indicate that the analytical models adapted from the welding process literature can capture the variation of the ripple widths and their wavelengths with process parameters.

The remainder of this paper is organized as follows. Section 2 presents a concise literature review of the microstructure studies on DED-printed 316L samples; Section 3 describes the experimental study to accumulate the various microstructure images; the multi-scale microstructure patterns of the DED-printed 316L samples from these experiments are presented in Section 4; concluding remarks are summarized in Section 5.

## 2. BACKGROUND AND LITERATURE REVIEW

In recent years, there has been a growing interest in studying the microstructure of DED-printed 316L components as they are being considered for multiple structural and biomedical component applications. Majumdar *et al*. [15] studied the microstructure of samples fabricated under different settings of process parameters, such as laser power density, scan speed, and powder feed rate, and determined the optimum processing conditions to print 316L with a homogeneous microstructure. They cataloged the different grain clusters occurring in different parts of the sample. They also noted that equiaxed grains formed at the top layer due to uniform heat distribution and heat conduction to the neighbor layers. They found that when the scan speed increased from 2.5 mm/sec to 7 mm/sec the area fraction of porosity decreased from 15% to 2% at a laser power density of 0.091 kW/mm$^2$. This was because high scan speeds induce sharp thermal gradients and solidification rates, reducing gas entrapment and hence micro-porosity. Increasing the scan speed from 2.5 mm/sec to 5mm/sec also reduced the range of grain sizes from 6-12 µm to 6-9 µm at a power density of 0.031 kW/mm$^2$. The powder feed rate had a negligible effect on grain size.

Zhang *et al*. [16] studied the influence of three process parameters, namely laser power, scan speed, and powder flow rate on the microstructure and mechanical properties of 316L cladding. As scan speed increased from 2 mm/sec to 10 mm/sec the secondary dendritic arm spacing (SDAS) changes from 7 µm to 2 µm, resulting in fine grains. It was concluded in the study that the scan speed was predominant in determining the microstructure along the build direction. Farshidianfar *et al*. [17] studied the effect of the cooling rate and melt pool temperature (monitored in real-time) on samples produced under high laser power (1.1kW IPG fiber laser YLR-1000-IC). The microstructure formed under a high cooling rate (764° C/s) reduced the grain size from an average grain diameter of 15.9 µm to 5.6 µm and thereby increased the grains per unit area from 3,968/ mm$^2$ to 31,744/ mm$^2$. They also discovered that scan speed has a linear relationship with cooling rate and melt pool temperature, with an increasing trend with cooling rate and a decreasing trend with melt pool temperature. Weng *et al*. [18] fabricated six 316L rods of diameter ranging over 1.16-1.36 mm and lengths ranging over 14.3-15.5 mm under different settings of laser power, powder feed rate, and scan speed. They observed that the powder feed rate influences microstructure. High powder feed rates (2.81 g/min) and scan speeds (0.6 mm/sec), promote rapid cooling, resulting in fine microstructures. The outer region of the rod was composed of an austenite phase and the inner region revealed the duplex phase (austenite and ferrite). This was because the outer wall dissipates heat in the convection mode which resulted in a higher cooling rate thus generating the austenite phase. Yang *et al.* [19] investigated the impact of laser power (400 W, 600 W, 800 W, 1000 W) on the microstructure of 316L samples using Scanning Electron Microscopy (SEM) and Electron Backscatter Diffraction (EBSD). They discovered that all the samples exhibit a cellular structure and that the size of the cellular structure grew with laser intensity, ranging from 7-14, 12, and 22 µm, respectively. Yang *et al.* [20] observed the microstructure of 316L samples produced at four different scan speeds (i.e., 6, 8, 10, and 12 mm/s) with constant laser power (1 kW). They claimed that when the scan speed increases, the grain size reduces while the porosity increases.

Boisselie and Sankare's study [21] employing coaxial laser cladding nozzles patented by IREPA LASER [22] indicates that the powders with low porosity caused poor powder flowability (quantified in terms of the powder flow rate through an orifice) and macroscopic defects in the parts. Mahamood and Pinkerton [23] inferred the microstructure based on different sizes of powder and metal shavings and measured the SDAS using SEM images at the mid-section of the printed part. It was observed that a powder particle size < 150 µm produced parts with a deposition efficiency (deposition rate / mass flow rate) of 49%-57%. Comparatively, larger powder sizes and shavings reduce the deposition efficiency to 23%-28%. The SDAS increased with laser power and decreased with the mass flow. In other words, the higher the energy input, the slower the solidification rate and the larger the SDAS. Wang *et al.* [24] studied the effect of $Cr_3C_2$ under the varying chemical composition of 316L powder such as (100% 316L, 95% 316L + 5% $Cr_3C_2$, 85% 316L + 15% $Cr_3C_2$, 75% 316L + 25% $Cr_3C_2$). They found that the columnar dendritic structure was predominant on the surface. They also studied the effect of the weight percentage of $Cr_3C_2$ and observed that an increase in the weight percentage of $Cr_3C_2$ produced a finer microstructure. And due to the dissolution of $Cr_3C_2$, fine carbides were present in the inter-dendritic eutectic phase.

Wang *et al*. [25] observed deposition of SS316L on a 316L substrate using a diode laser system at 1000 W and produced oblique thin wall parts ($\theta = 0°, 20°, 45°$) with two different strategies: (1) the part was produced by maintaining the scan parallel to the substrate and the oblique angle was achieved by decreasing the scan length uniformly. (2) the oblique angle was achieved by increasing z-increment steadily from the low side (0.1 mm) to the high side (0.2 mm) in every deposition layer. It was observed that with the increase in oblique angle, the solidification rate decreases, which resulted in a coarse grain microstructure. Guo *et.al*. [26] observed the deposition of 316L on a substrate in two orthogonal build directions, i.e., 0° and 90°. They noted that higher laser power produces high thermal gradients in the molten pool in the building direction of 90°. As the molten pool advances, the heat starts dissipating among previously deposited layers which eventually resulted in a lower cooling rate. This results in the formation of large dendritic grains in the direction of the thermal gradient. A higher cooling rate was achieved in the building direction of 0° which resulted in the formation of the austenite phase and produced a more homogenous microstructure. Along similar lines, Mukherjee [27] studied the effect of build geometry on microstructure by manufacturing two types of specimens. One built along the direction (i.e., scan direction) parallel to the longest side and another perpendicular to the longest side. Their study suggested that (a) columnar grains on both the specimen, (b) the manufacturing cycle time for the second specimen was larger, and (c) finer microstructure existed on the second specimen with some interlayer porosity (due to the high cooling rate).

Khodabakhshi *et al*. [28] studied the change in microstructure along the build direction and found that the first few layers and the topmost layer are majorly comprised of equiaxed grains because of rapid solidification around the edges. However, during the deposition of the subsequent layer, intermediate layers experienced re-heating and dilution, resulting in a final columnar dendritic grain structure. Feenstra *et al*. [29] studied the microstructure evolution along the build direction and observed stacked patterns of the columnar structure near the substrate. The study also showed the different patterns of columnar grains emerging at different sections in the build direction. Since the laser scan was bi-directional (i.e., back and forth raster scan pattern), the columnar grains formed a zig-zag pattern in the middle of the sample, and large columnar grains aligned along the build direction near the top surface. Balit *et al*. [30], [31] manufactured 316L specimens under two different build orientations at two vertical increments (i.e., the spacing between consecutive layers) of 0.2 to 0.12 mm. Their subsequent SEM and EBSD study suggest that the grains were preferably oriented along the build direction. They calculated the different grain size statistics (aspect ratio, grain diameter, number of grains, surface occupied), and found that small grains of average diameter $8.5 \pm 3.5$ µm (per their categorization) was present between the layers and large grains present over the layers. Zhi En *et al.* [32] considered three scan strategies, namely long-unidirectional, short-unidirectional, and bidirectional. It was also found that the mean grain size was greatest in short unidirectional (15.7 µm) and smallest in long unidirectional (9.7 µm), with the difference in grain sizes due to the higher temperature at the initiation point of each consecutive short unidirectional raster scan path with the shorter length of each raster scan path and the heat generated from the previous deposition path, resulting in a lower thermal gradient. The temperature at the initiation point of each long unidirectional raster scan path was lower due to the comparatively longer raster scan path, resulting in a steeper thermal gradient. The needle-like microstructure and acicular crystallite phase were observed in all the specimens.

Saboori *et al*. [33] studied the Primary Cellular Arm Spacing (PCAS) [34] to analyze the effect of the cooling rate at different heights from the substrate. It was observed that PCAS increased from 2.8 µm to 4.6 µm with the increase in height of the substrate from 2 mm to 14 mm because of the slow cooling rate. The relatively higher cooling rates at the top layer resulted in a low PCAS value of 3.3 µm. The microstructure analysis revealed a duplex phase and found out the percentage of delta ferrite decreased with the increase in the distance from the substrate up to 14 mm. Wang [35] deposited 316L between two horizontal magnets, in the presence of an external magnetic field and discussed how it affected the microstructure using SEM and XDS results. It was observed that an increase in the magnetic field from 0 T to 1.8 T changed the morphology of the ferrite phase and increased the cellular dendritic spacing. It showed a major effect on Cr and Ni content present in 0 T versus 1.8 T samples; Cr content increases immensely in the ferrite phase and Ni content increases slightly in the austenite phase, respectively. Zheng et al. [36] deposited 316L samples using a Yb fiber laser and observed the columnar grain structure due to epitaxial growth from the prior solid interface within each layer. The cellular colonies were identified on the surface of each layer.

Huang *et al*. [37] compared the 316L and Inconel 625 built under the different process parameters and discussed the relation between solidification parameters, thermal characteristics, and process parameters. They observed that the combination of high scan speed and low laser power generated a finer microstructure. It was found that product G×R, where G (°C/mm) refers to thermal gradient and R (mm/sec) refers to solidification growth rate increased from bottom to top of the clad and found that the dendritic arm spacing was more sensitive to scan speed than the laser power. Rankouhi *et al*. [38] compared the parts manufactured from DED and Selective Laser Melting (SLM) and studied different characteristics. Grain structures produced in both processes were studied and concluded that in SLM there were a greater number of elongated grains that were aligned in the direction of the scan named as "mosaic" structure. In DED, the grains were elongated in a random direction and made 20-30° with scan direction. The aspect ratio of grains was higher in the DED process i.e., 3.09 than in SLM i.e., 2.07 in the scanning plane.

A few studies focused on morphology and microstructure resulting from a functional grading of 316L with other materials. Nie *et al.* [39] deposited 316L and SS431 and found as the equiaxed grain structure dominates in the zone of 316L until 49% (wt.%) and turns into a mixture of columnar and equiaxed between 72-100% (wt.%). Zhang *et al.* [40] deposited 316L with IN 625 and concluded that in both the samples columnar grains were observed near the interface and grow along with the laser scan direction. Banait *et al.* [41] fabricated the FGM of Ni-Cr-B-Si and 316L and compared the heat-treated and as-built sample and concluded the epitaxial growth of dendrites in 316L zone in the as-built sample and coarse recrystallized grains in the heat-treated sample and it was also observed that dendritic structure dominates when wt.% of 316L crosses 50%. Kim *et al.* [42] deposited 316L and P21 steel and studied morphology along the built direction using EBSD. It was concluded that 316L layers were equipped with fully austenitic coarse grains and were heterogeneous because of high heat input. Li *et al.* [43] manufactured 316L and IN 718 using thermal milling on Hybrid DED and characterized microstructure and phase evolution and found that element segregation was absent in 316L, and both the metals consisted of the austenite phase. Zhang *et al*. [44] studied the microstructure of copper-stainless steel hybrid components along the build direction. It was found that the upper layers of stainless steel consist of equiaxed grains because the G/R ratio keeps decreasing along the build direction and due to the high G/R ratio near to the substrate the columnar dendritic grains were present.

A few computational studies peripherally discussed certain physical mechanisms that are relevant to ripple formation. One such study by Haley *et al.* [45] investigated the particle impact on the melt pool and wettability effects in DED and found that the impact velocity of the powder particle in the melt pool causes ripple formation in the molten phase. Aggarwal *et al.* [46] discussed the role of impinging powder particles in the melt pool and found that the marangoni convection controls the shape of the melt pool during the heating stage, and the impact velocity of the powder particle creates the ripples during the deposition stage. Senin *et al.* [47], [48] mentioned the ripples in powder bed fusion as surface texture features. Tang *et al.* [49] investigated the formation of humps and ripples in 316L produced using SLM. The keyholing phenomena, which caused the molten pool to oscillate, was credited for the ripple's formation. The molten liquid was forced backwards, causing a tailing event that resulted in the formation of ripples. They also compared the simulation results for melt pool width and depth to the experimental data, finding a 12 percent difference between the two. Yang *et al.* [50] reported a formation of "coagulation ripples" on the surface of AlSi10Mg fabricated by SLM. The ripple formation was caused by marangoni convection. They have also discovered that as the linear energy density (LED) rises from 4.93 J/cm to 14.8 J/cm, the ripple structure changes from a tapered to a circular pattern. Hu *et al.* [51] developed a mesoscopic model that replicated the transport phenomenon of SLM build 316L under varying ambient pressure. They discovered that ripples on the surface have an impact on the surface roughness of the as-printed tracks. The oscillation of the molten pool during keyholing was discovered to be a crucial factor in the creation of ripples. They discovered that as the pressure drops from 100 Pa to 1 atm, the ripples get smaller (height = 1 μm, spacing = 25 μm) and the surface becomes smooth.

Key differences between SLM and DED on ripple formation include that the SLM process has low energy input, resulting in smaller or shorter wavelength ripples. In contrast, the DED is a high-energy process due to which many different phenomena like marangoni convection, recoil pressure, laser interactions lead to formation of ripples, and it was discovered that laser power was a critical factor influencing the geometry of ripples, such as their curvature (circular or elliptical) and wavelength.

Apart from the differences mentioned above, these studies, however, overlook the morphological signatures of the ripples, as well as an understanding of how ripples inform the microstructure along scan direction and the influence of process parameters on ripple geometry and the resulting microstructures.

In summary, out of over 50 works studied the microstructure of DED-printed parts, 40 employed optical microscopy. Among these, 8 focused on the effects of different printing (build) strategies [25]–[32], 20 investigated the effect of different process parameters [5], [9], [12], [15], [17]–[19], [23], [26], [28], [30], [36]–[38], [52]–[56], and 4 considered the effect of powder compositions [21], [23], [24], [57]. Three of them calculated dendritic arm spacing to understand the solidification and cooling rate during the process, and 36 studies discussed the grain formation at melt pool fronts along the build [5], [10], [15]–[34], [36], [37], [39], [40], [52]–[56], [58]–[66]. Only 5 studies discussed the structure along scan direction [12], [26], [27], [34], [61]. Additionally, 8 reported additional SEM studies to assess the percentage of ferrite and austenite formations, and calculated the Cr/Ni equivalent ratio [24], [33], [58], [65], [67]–[70], and 7 EBSD studies characterized the crystallographic grain structure [30], [35], [38], [42], [71]–[73].

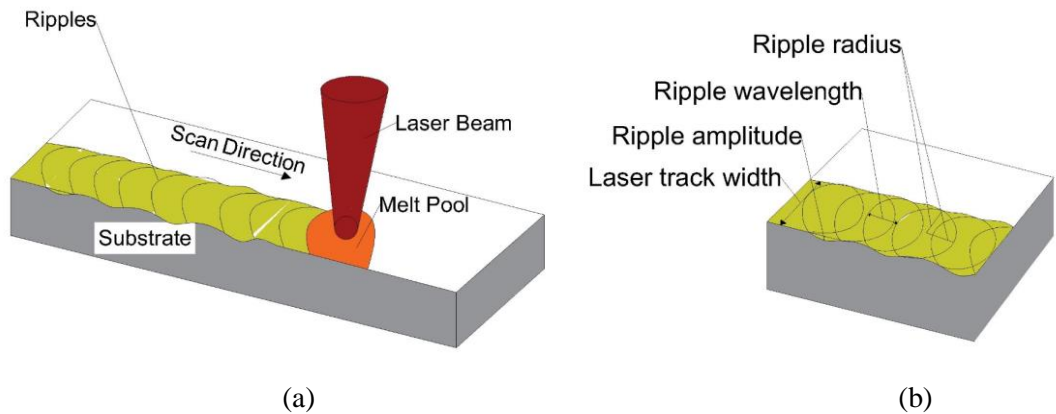

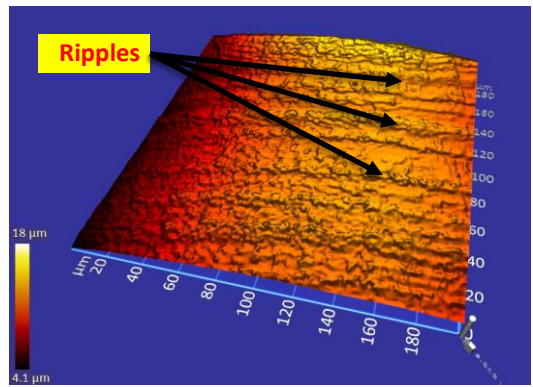

(c)

Fig. 2: (a) schematic of ripple formation on the surface, (b) ripple nomenclature, (c) ripples observed on the as-printed surface under white light interferometry.

More pertinently, none of the earlier micrography studies addressed the ripple formations that are commonly observed in DED processes (see Fig. 1b). Their physical underpinnings are studied in welding processes [74], [75] but have been overlooked in the context of DED. Some recent studies even attributed the ripple signatures to the melt pool geometry.

The experimental studies reported in the sequel suggests that the ripples in DED exhibit prominent near-harmonic wave patterns. For example, a representative heightmap of the surface of a DED-printed component obtained using white light interferometry (Fig. 2c) captures variations of amplitudes and wavelengths (both of the order of 10 µm). Our experimental study also reveals that the process parameters, especially the laser power, have a systematic influence on the ripple formations as well as the scan track marks (Fig. 2). The morphological formations of the track marks and the ripples critically inform the microstructure clusters under different magnifications. The present experimental and optical micrography study focuses on the direct observation of these formations, which have been largely ignored in the prior literature.

## 3. EXPERIMENTAL APPROACH

### 3.1 Experimental setup

The experiments consisted of printing stainless steel 316L samples of dimensions 10 mm × 10 mm × 10 mm using an Optomec-LENS® MTS 500. The machine tool is essentially a 4-axis CNC machine that automatically controls the worktable motion in X and Y directions, motion of the laser and a milling head along a Z (vertical) axis, and the rotation of a horizontal spindle, which can be used to clad and repair a variety of freeform parts. This machine tool uses an IPG YLR-1000 fiber laser with a spot size of 600 µm and a wavelength of 1070 nm. It also comprises two powder feeder hoppers with an option to extend to four hoppers. The powder from individual hoppers is transported to a mixing chamber and finally delivered through four nozzles using pressurized argon. Besides powder transport and delivery, argon also helps to reduce oxidation during deposition and serves as a shielding gas during deposition. The machine tool is capable of maintaining oxygen levels in the chamber to less than 40 ppm. For the current set of experiments, an open atmosphere (OA) mode was used. The schematic and working principles of DED are illustrated in section 1, Fig 1. A highly focused laser beam interacts with the powder particles and forms a melt pool which eventually solidifies along the direction of the scan. The machine, through a Siemens 840D controller, provides control over multiple process parameters such as powder composition, powder feed rate, laser power, hatch spacing, scan speed, and dwell time during the process.

### 3.2 Experimental design

As noted in Section 2, laser power, scan speed, and dwell time are known to significantly affect the cooling and heating rates, and thereby the scan track marks and ripple formations. Also, as elaborated in Section 4, these process parameters and the formations have a significant bearing on the microstructures in the part. We therefore focused on how these formations and microstructures of DED-printed 316L at different scales vary with these process parameters,

The settings of these three process parameters were chosen according to an orthogonal Latin hypercube experimental design [76]. This experimental design helps to achieve a maximal "space-filling" exploration of the parameter space spanned by the laser power, scan speed, and the dwell time. We also considered the limits (mostly the upper bounds) on the process parameter combinations at which we can operate the machine safely. For example, the maximum laser power should not exceed 600 W, as higher laser power settings tend to create fumes within the OA operating environment. The maximum scan speed was set to 8.46 mm/sec to ensure proper melting. The maximum dwell time was set to 40 s, beyond which the part cools down significantly between successive scans, causing surface distortions [77] that disturb the focusing distance of the laser. Within the set bounds of the process parameters, nine experimental points, as summarized in Table 1 were selected, i.e., one cubic sample was printed under each of the nine conditions.

Table 1 DED process parameters settings used to print the nine samples for micrographic studies

| Experiment | Laser power (w) | Scan speed (mm/sec) | Dwell time (sec) | Linear Energy Density (J/mm) |
|---|---|---|---|---|
| 1 | 487.5 | 6.56 | 35 | 74.31 |
| 2 | 525 | 3.70 | 0 | 141.9 |
| 3 | 562.5 | 8.46 | 15 | 66.48 |
| 4 | 600 | 1.80 | 30 | 333.33 |
| 5 | 450 | 4.65 | 20 | 96.77 |
| 6 | 412.5 | 2.75 | 5 | 150 |
| 7 | 375 | 5.60 | 40 | 66.96 |
| 8 | 337 | 0.84 | 25 | 401.19 |
| 9 | 300 | 7.40 | 10 | 40.54 |

### 3.3 Experimentation and sample preparation procedures

AISI Stainless Steel 316L with powder particles ranging between 44-106 μm in diameter was used in this study. Pertinently, 316L is one of the most popular grades of stainless-steel owing to its exceptional combination of strength and corrosion resistance properties, and is widely used in many engineering domains, including nuclear, aerospace, and automotive sectors. The elemental composition of the powder is mentioned in Table 2. The DED process involved printing nine cubic 316L samples, at process parameter settings stated in Table 1, onto a rectangular 316L substrate of dimensions 76 mm × 76 mm that has an identical chemical composition as the powders. All samples were printed under the same environmental condition and with the same powder feed rate (feeder RPM: 5 RPM). After the deposition, the top surface of printed parts was machined (end-milled) as-printed, i.e., the substrate surface on which all the samples were built served as a datum and the machined surface was therefore flat within the machine precision relative to the scan direction. As an exception, however, the top surface of as-printed sample 4 was not parallel to the scanning plane due to the high laser power input. Consequently, post-milling, the resulting micrographs capture the morphological and microstructural features of multiple layers along the scan direction.

Table 2 Chemical composition of SS316L powder

| Element | Cr | Ni | Mo | Mn | Cu | P | Si | C | S | Fe |
|---|---|---|---|---|---|---|---|---|---|---|
| Wt.% | 18 | 13 | 2.5 | 2.0 | 0.5 | 0.025 | 0.75 | 0.03 | 0.01 | Bal. |

Subsequent to machining, the samples were detached from the substrate using a bandsaw and prepared for microstructure analysis. All the samples were mounted in an epoxy resin with the top surface exposed for polishing. The samples were polished according to the usually metallurgical sample preparation for microstructure study. The polishing was performed on a Buehler Automet 250 polisher in multiple steps. The initial steps were conducted at 100 rpm, using 800, 1000, and 1200 grade emery pads, respectively, and the final finish polishing was conducted at 50 rpm using a micro-cloth and colloidal silica. The polished surfaces were etched for 90 seconds in an Aqua Regia etchant (which contains 3:1 part of HCL and $HNO_3$). The morphology and the microstructure were observed on the Olympus BX 51 microscope at four different magnifications – 5x (500 µm length scale), 10x (200 µm length scale), 20x (100 µm length scale), and 50x (50 µm length scale).

## 4. RESULTS FROM OPTICAL MICROGRAPHY STUDIES

As noted in Section 2, a vast majority of earlier micrography studies of DED-printed 316L had focused on the build direction. A very few discussed the microstructural patterns along the laser scan direction. Studying the micrographs along the scan direction can contribute to the understanding of the solidification process, the surface morphology, microstructure, and certain defects [77].

Also, as noted earlier, the present study examines the microstructure patterns from the optical micrographs (OMs) taken at four different magnifications. Since the same powder feed rate was employed for all the samples, the deposition rate depends only on the scan speed and the melting process. The study captures the effect of the laser power and scan speed on the ripple formation, as well as laser scan track marks and void formations in the scan direction, and how they inform the heterogeneity of the microstructure.

**Observations at 500 µm**: Fig. 3 shows the representative micrographs which were obtained at 5x magnification, i.e., the microstructures over a field of view of 5.3 mm (500 µm length scale). Fig. 3(a) was obtained from sample 4 that was printed at 600 W laser power and 1.80 mm/sec scan speed. Fig. 3(b) was obtained from sample 5 that was printed at 450 W laser power, and 4.65 mm/sec Fig. 3(c) was obtained from sample 9 that was printed at 300 W laser power and 7.40 mm/sec scan speed.

It may be noted that it is hard to observe the detailed microstructure at such low magnification. However, the large fields of view at this scale allow us to capture the patterns of the laser track [78] as well as the ripples. All three micrographs in the figure show the laser track marks consisting of ripple formations (circular arcs with different color shades) on the trailing edge of the laser head along the scan direction [79]. The various shades present in micrographs are due to the differential action (selectivity) of the etchant on various chemical compositions and molecular structures present at the sample surface [80].

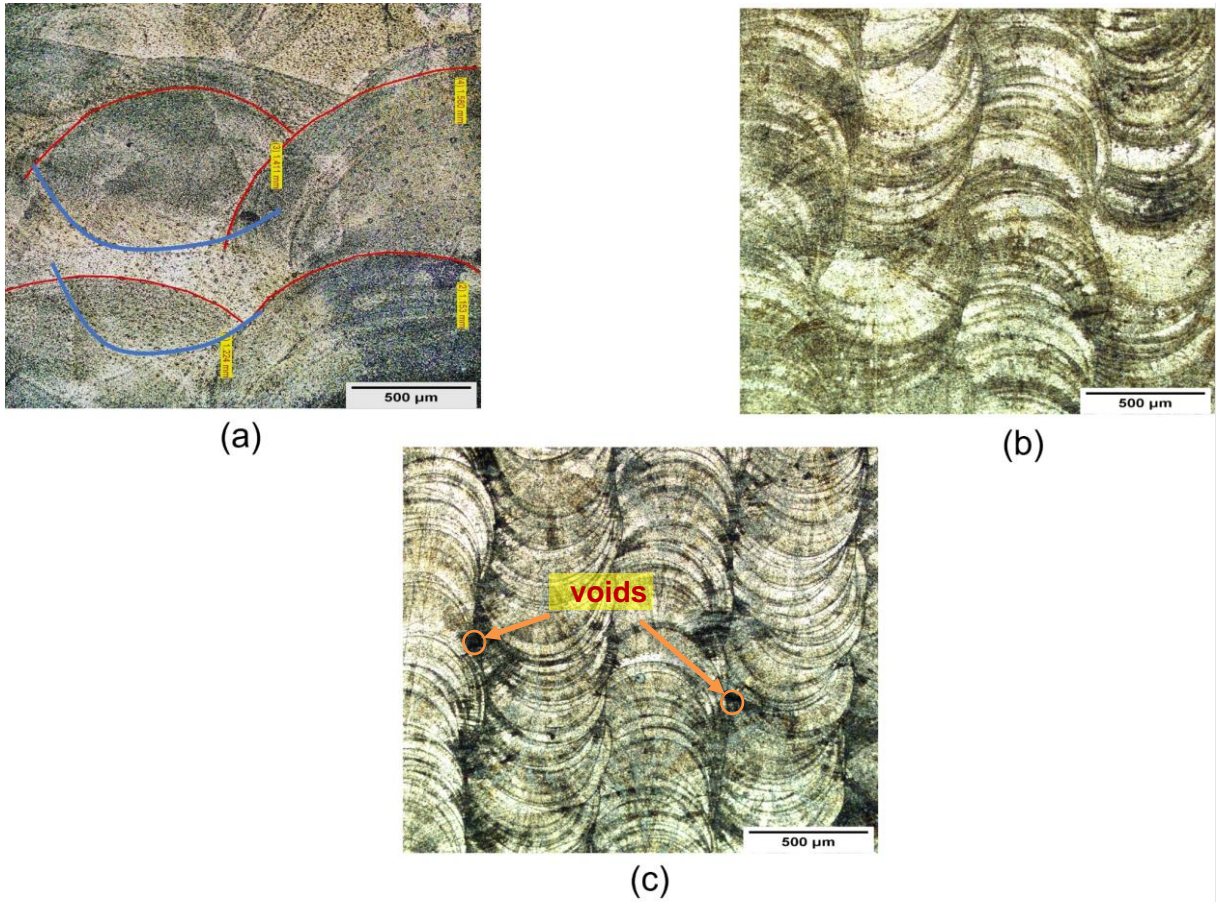

Fig. 3: Optical micrographs at 500 μm scale for laser power (a) 600 W (b) 450 W (c) 300 W

We further measured the laser track width and the ripple curvature along the scan direction. We observe that the curvature was uniform across the perimeter of every ripple. The ripple curvature (for our case ripple radius) was calculated using a 3-point arc estimate (a utility available in AutoCAD). The average radius of curvature (ROC) was estimated for five ripples on each laser track and their distribution, and the average statistics are presented. As evident in Fig. 4(a-d), the dimensions of the laser track majorly depend on the laser power and to an extent on the scan speed. The ROC appears to increase almost linearly with the laser power. The other variations, although statistically significant, do not conform to a specific pattern, and may benefit from additional investigations.

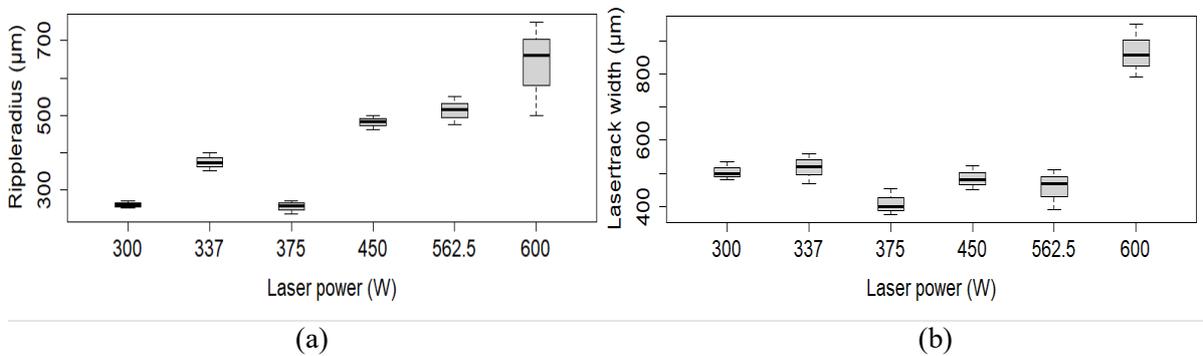

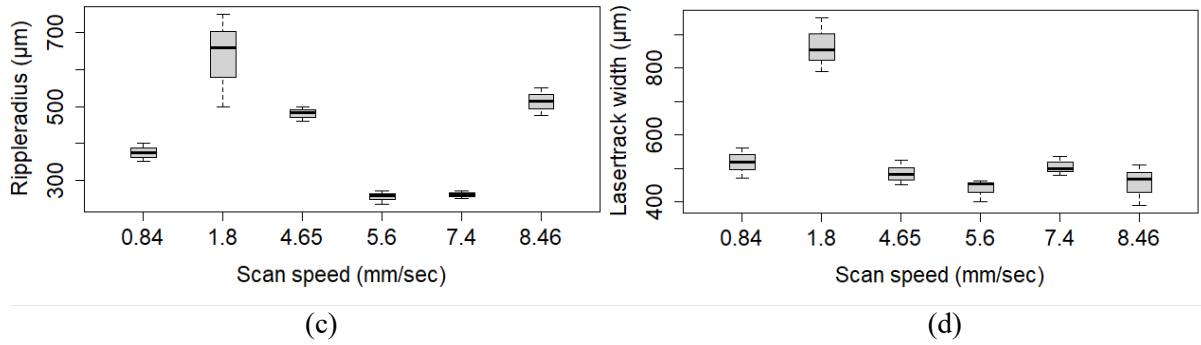

(c)　　　　　　　　　　　　　　　　(d)

Fig. 4: Variation of the (a) ripple curvature and (b) laser track width with laser power, and (c) ripple curvature and (d) laser track width with scan speed.

The average laser track width calculated for sample 4 (600 W, 1.8 mm/s) was found to be 855 μm which is near twice that of sample 9 (300 W, 7.4 mm/s) i.e., 467 μm. Sample 5 (450 W, 4.65 mm/s) shows an overlapped laser track with an average laser track width of 550 μm. Fig. 4(b) shows the variation of the laser track width with the laser power. It can be inferred that a low scan speed and high laser power results in wider laser tracks.

These patterns are consistent with the earlier studies on the variation of the clad-width or single laser track with the change in scan speed and laser power [81]. With the laser spot size kept constant, low scan speeds and high laser power lead to high laser energy density, which increases the volume and the temperatures of the melt pools [17]. More importantly, the surface tension of molten 316L decreases with increasing temperature, which increases the width of the melt pool with temperature [75].

Also, a laser track experiences complicated thermal cycles near the track boundaries—this includes altering the heat transfer patterns due to the effect of the temperatures of the previously laid adjacent tracks, and remelting of solidified tracks at the boundaries with subsequently laid tracks [82]. This introduces interesting morphological formations such as voids as seen in Fig. 3(c) and determines the dendritic structure observed at higher magnifications. Ripples with wavelength < 0.508 mm (as in sample 9 shown in Fig. 3(c)) emerge under high scan speed and low laser power. The ripple wavelength increases to 1.812 mm when scan speed is reduced (as in sample 5).

**Observations at 200 μm:** Fig. 5 shows the representative optical micrographs obtained at 10x magnification. Here, the field of view was 2.65 mm. Fig. 5(a) was obtained from sample 4 that was printed at 600 W laser power and 1.80 mm/sec scan speed. Fig. 5(b) was obtained from sample 5 that was printed at 450 W laser power, and 4.65 mm/sec scan speed. Fig. 5(c) was obtained from sample 9 that was printed at 300 W laser power and 7.40 mm/sec scan speed.

The micrographs capture the spatial distribution of different microstructure clusters relative to the laser track boundary and the ripples. Fig. 5(a) is taken from the (heat-affected) zone adjoining two laser tracks. The structure is mostly composed of columnar dendrites with equiaxed grains present inside the laser track. At lower power settings (Figs. 5(b, c)), the dendritic structure is not apparent as the grains tend to be highly refined.

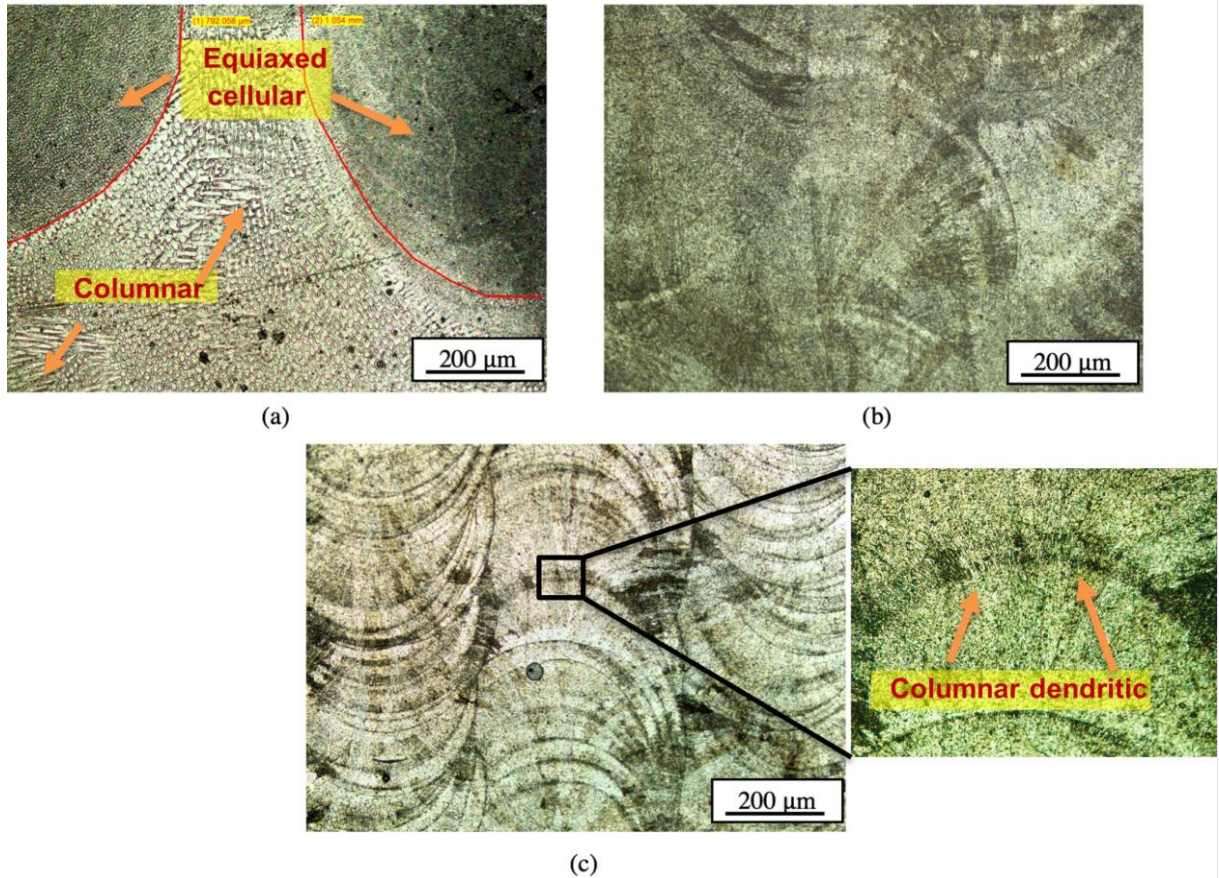

Fig. 5: Optical micrographs at 200μm scale for laser power (a) 600 W (b) 450 W (c) 300 W

The microstructure formations in a DED process can be explained in terms of the ratio of the local temperature gradient G (°C/mm) and solidification growth rate R (mm/s), i.e., the ratio G/R [83]. Due to a large distribution of thermal gradients in the melt pool, we are more likely to find heterogeneous microstructure as seen in Fig. 5(a) [68]. Evidently, the thermal gradients are the highest radially from the center to the edge of a ripple, as well as from the center to the edge of a scan track (i.e., along the width of the track). Consequently, the solid-liquid interface advances preferentially along the normal to the ripple boundaries and across the track. The resulting microstructure consists of equiaxed cellular grains at and outside a ripple boundary and in the interior of a ripple, and columnar dendritic structure oriented almost radially from the ripple boundary into the interior [69], [73].

**Observations at 100 μm:** Fig. 6 shows the representative micrographs at 20x magnification of a microscope field of view of 1.32 mm (at 100 μm length scale). Fig. 6(a) was obtained from sample 4 that was printed at 600 W laser power and 1.80 mm/sec scan speed. Fig. 6(b) was obtained from sample 5 that was printed at 450 W laser power, and 4.65 mm/sec. Fig. 6(c) was obtained from sample 9 that was printed at 300 W laser power and 7.40 mm/sec scan speed.

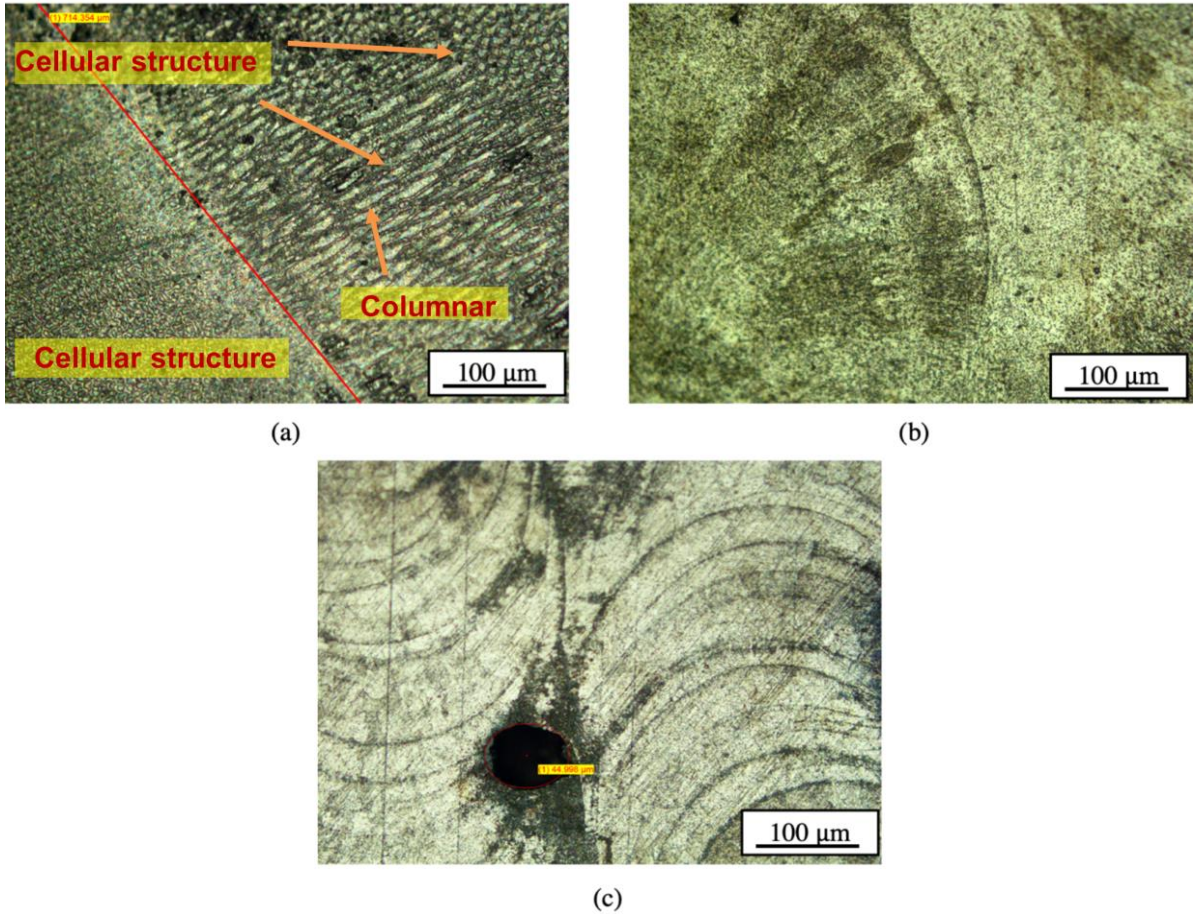

Fig. 6: Optical micrographs at 100µm scale for laser powers (a) 600W (b) 450W and (c) 300W

As shown in the figures, the spatial distribution of dendritic structures and their preferred orientations becomes more evident at this scale. For example, Fig. 6(a) is captured near the perimeter of a ripple shown in Fig. 5(a). It reveals that at 600 W (high laser power) the microstructure consists of long columnar dendrite and cellular network structures with a sharp boundary separating these. The cellular structure exists in a few columnar grains (marked in Fig. 6(a)) and has a cell length in the range of 0.06-0.1 mm. This cellular structure shows intragranular features which means the sub-grain structure (cellular structure) is constricted inside the large grains (columnar structure). The sharp cooling rates prevailing in DED give rise to smaller-sized cells (seen at 50 µm scale length) compared to in a laser welding process [84]. These structures are spread over an average area of 270 µm$^2$. At lower power settings, just as with 200 µm magnification microstructure, the dendritic structure is much finer and is not clearly visible at this magnification (see Fig. 6(b)). Nonetheless, the dendrites seem to have grown preferentially oriented radially inward or in the direction of a moving heat source (i.e., laser beam) from the perimeters of ripples.

At a lower power setting of 300 W and high scan speeds (7.40 mm/sec), defects such as voids become apparent on the surface (see Fig. 6(c)). These voids occur due to the uneven melting of the powder, especially near the junction of laser tracks (consequence of the low energy input at the specified condition), coupled with the lack of sufficient material to compensate for the shrinkage (as a result of high scan speeds). Additionally, an equiaxed cellular structure is observed near the center of the ripple because of the high cooling rate of nearly $10^3 - 10^4$ K/s and moderately high solidification rates [19].

**Observations at 50 µm:** Fig. 7 shows the representative micrographs at 50x magnification at microscope field of view of 0.53 mm (at 50 µm length scale). Fig. 7(a) was obtained from Sample 4 that was printed at 600 W laser power and 1.80 mm/sec scan speed. Fig. 7(b) was obtained from Sample 5 that was printed at 450 W laser power, and 4.65 mm/sec. Fig. 7(c) was obtained from Sample 9 that was printed at 300 W laser power and 7.40 mm/sec scan speed.

At this magnification, the micrographs show varied shades on a grey scale. Earlier studies suggested that the dark and light portions at this magnification are likely due to the differential etching among the various material species, especially the austenite (light) and ferrite (dark) phases obtained after solidification. Additionally, Fig. 7(a) shows a contrast between the cell core and boundary.

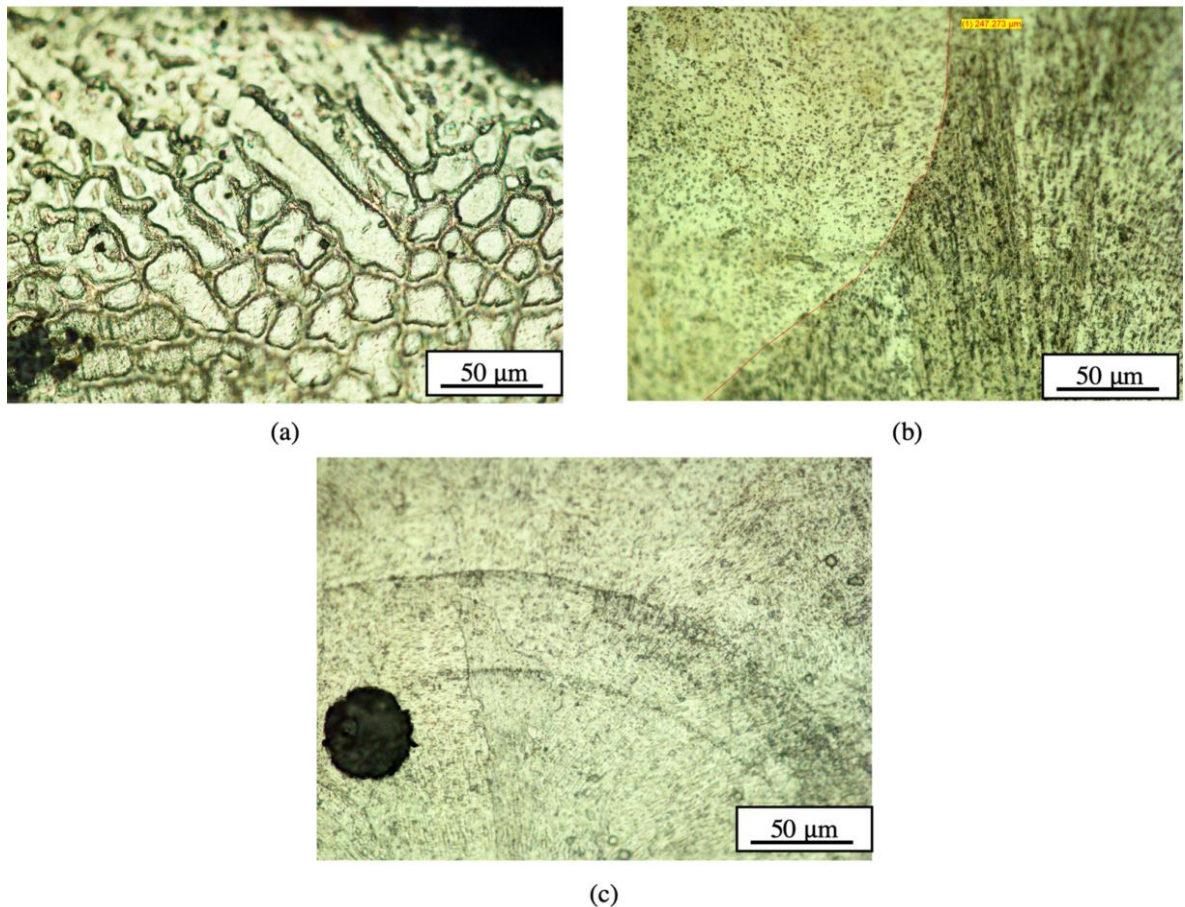

Fig. 7: Optical micrographs at 50µm scale for laser power (a) 600W (b) 450W (c) 300W

The etchant used here, aqua regia, is known to preferentially attack the δ-ferrite and carbides such as $Cr_{23}C_6$ and $Mo_{23}C_6$ and thereby helps to identify the presence of the heavier elements at the cell boundary as seen in Fig. 8(a). As observed in the (Energy Dispersive Spectroscopy) EDS line scan in Fig. 8 (b), Cr and Mo were found to be higher in mass percentage at the cell boundary than in the cell core. On the other hand, Ni concentrations were found to be higher in the cell core than at the cell boundary. An EDS spot analysis was performed at 6 different locations at the boundary and the interior. The Cr and Mo mass percentages at cell boundary, calculated from this analysis, were found to be 18.55% (standard deviation of 0.85%) and 4.45% (standard deviation of 0.64%), respectively, and 15.56% (standard deviation of 0.53%), and 3%, (standard deviation of 0.47%) respectively, at the cell core. This statistically significant depletion of Cr and Mo in the cell core is consistent with the earlier studies [85].

As a DED-printed 316L solidifies, austenite is the primary solid phase to emerge from the liquid melt pool. This is consistent with the earlier observations of austenite (γ-phase) forming at the core, and δ-ferrite between the dendrites as seen in Fig. 7(b) [86]. The dark portions in Figs. 7(b&c) are indicative of the dendrites containing δ-ferrite (which was revealed after the etching) forming near the laser track boundary. Also, fast-directional solidification and re-heating of a deposit formed along a laser track (owing to the subsequent deposition of material along adjacent laser tracks) can induce micro-segregation of γ and δ-phases within a dendrite.

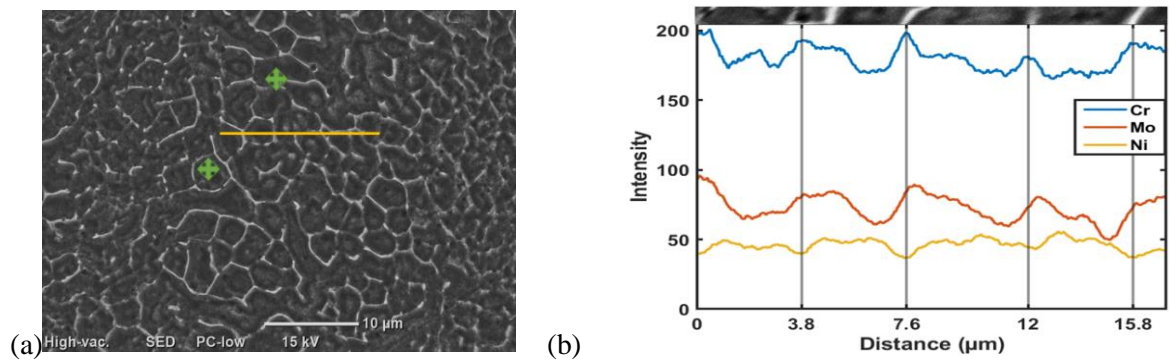

Fig. 8: (a) An Energy Dispersive Spectroscopy (EDS) image marked with two different spots, one at cell boundary and another at cell core, (b) An EDS line scan capturing Cr, Mo, and Ni variations.

## 5. DISCUSSION

### 5.1 Heterogeneity of the DED dendritic structure:

Fig. 9 summarizes the different microstructures, including the shape and the size of various formations at different positions on the sample surface, namely, the centre of a ripple, the perimeter of a ripple, and the edge of scan tracks. As noted earlier, the microstructure depends on the spatiotemporal distribution of the temperature gradient G (°C/mm) and solidification growth rate R (mm/s). While the ratio G/R, controls the solidification mode, the product G×R decides the size of grains [83]. Moving radially inward from the fusion line (or near the outer edge of ripple), the structure changes from equiaxed to columnar dendritic to cellular as shown earlier in Fig. 5(a). At a low laser power (300-450 W) and high scan speeds (4.65-8.46 mm/sec), a fine grain structure was observed (see Fig. 10). At high laser power (> 450 W) and low scan speeds (0-4.65 mm/sec) a coarse dendritic structure was apparent. In addition to the size of structures, the laser power density has a major effect on the formation of ripples (see Fig. 4).

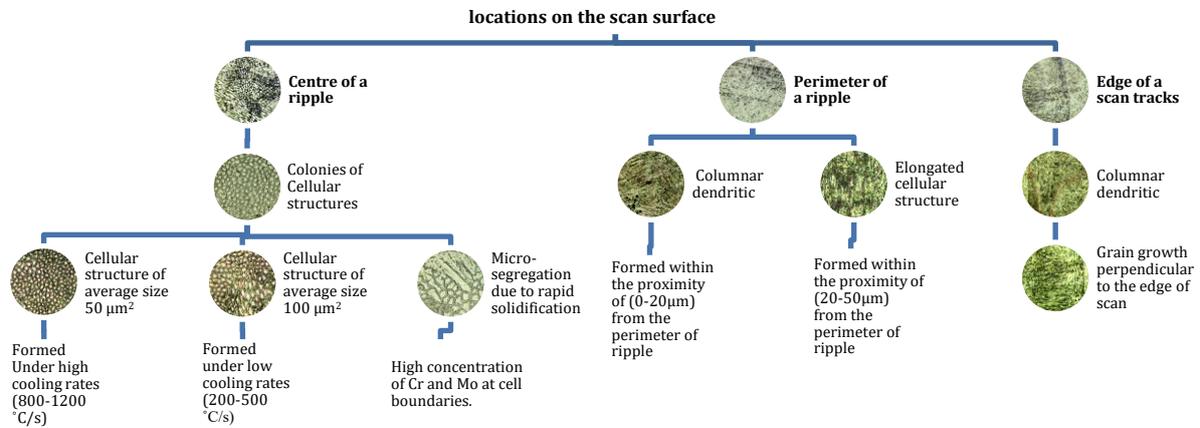

Fig. 9: The salient structures observed across the various portions of the scan surface of a DED-printed 316L samples.

Table 3 Different segments classified from Microstructure Hierarchy and mapped in solidification map

| | |
|---|---|
| 1 | Centre of ripple |
| 1a | Cellular structure of size 50μm² at center of ripple |
| 1b | Cellular structure of size 100μm² at center of ripple |
| 1c | Segregated cellular structure |
| 2 | Perimeter of ripple |
| 2a | Columnar dendritic structure between 0-20 μm from the perimeter of ripple |
| 2b | Elongated cellular structure between 20-50 μm from the perimeter of ripple |
| 3 | Edge of the scan track |
| 3a | Columnar dendritic structure |

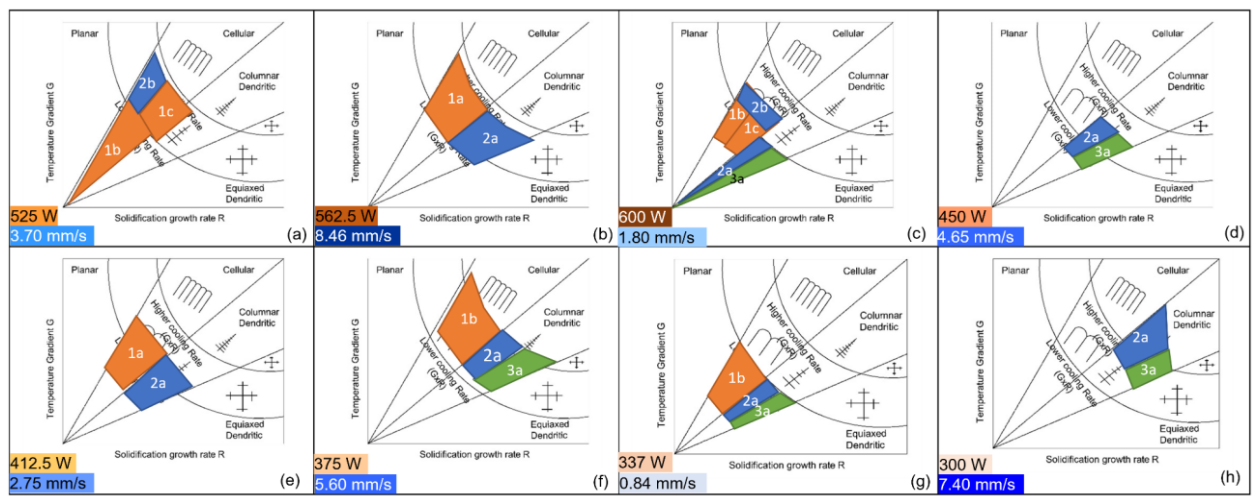

Fig. 10: A solidification map showing the different areas marked over specific grain structures observed at different locations on the surface.

## 5.2 Effect of Laser Power and Scan Speed on Ripple Geometry and Laser Track Width

The ripple formations, commonly observed in the welding processes, are attributed to five different phenomena, namely, thermocapillary force, power source variations, periodic swelling in the liquid metal, flow instability (recoil effect) due to evaporation, and solidification dynamics [75]. Cheever *et al.* [87] observed the surface ripples on different thin metal sheets in gas tungsten arc, spot, and seam welding processes. They observed that sharp variations in the plasma pressure caused due to intermittent shut-off of the arc cause the oscillation of the weld pool during solidification and create ripples on the surface. Anthony and Cline [74] studied the surface ripples using laser surface melting techniques used for enhancing the surface property like transformation hardening of the surface. They inferred that the surface tension gradients were the main cause behind the surface ripples. Wie *et al.* [88] studied the surface ripples in Electron Beam Welding of Al 1100 and SS304 specimens. They observed that the average amplitude and pitch of ripples increased with increasing dimensionless beam power, Marangoni, Prandtl, and Biot numbers and decreasing Stefan, and Peclet numbers. Wie *et al.* [75] studied the temperature dependence of surface tension (gradient of surface tension relative to temperature) in electron beam welding and noted that in low S concentration alloys, the surface tension gradient is negative leading to the formation of the shallow and wide melt pool, and positive in high S concentration alloys, leading to deep and narrow melt pools. Wie [89] observed that ripple roughness increases with the decreasing welding (scan) speeds. Based on these prior findings reported on the welding process, the ripple formation in the DED process may be attributed to the following two phenomena.

*1) Recoil Pressure*: During the melting process under the laser beam in DED, the particles move within the liquid metal with a certain kinetic energy. This kinetic energy increases with temperature and when these particles reach near the top surface of the melt pool, they have enough energy to escape the liquid metal into vapor. As shown in Fig. 11, vapor particles eject with momentum ($m_v v_v$) from the liquid surface, causing the liquid to recoil with a momentum of ($mv$) where $m$ is the mass of liquid surface and $v$ is the velocity of the liquid surface. This exchange of momentum at the surface creates a recoil pressure which acts on the melt pool surface against the force induced by the vapors. It depresses the molten layer downwards and disperses the liquid to the edges of the melt pool.

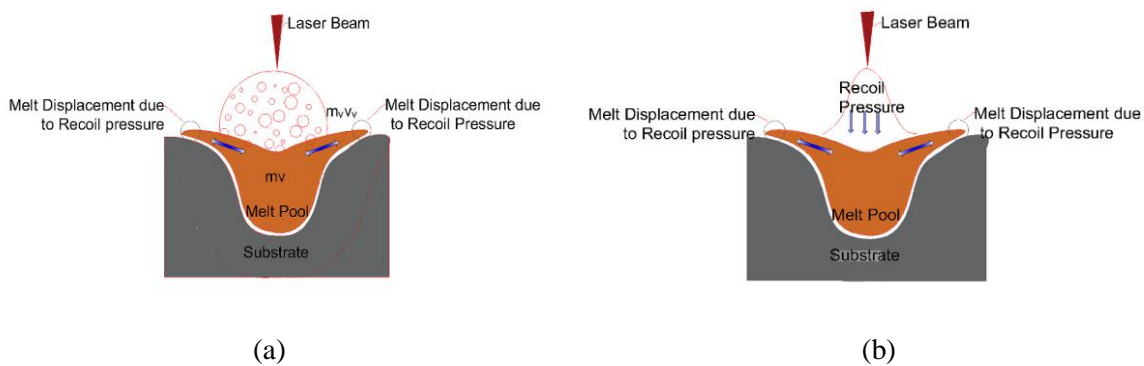

(a)                                         (b)

Fig. 11: (a) Exchange of momentum induces the recoil pressure, (b) Liquid metal is displaced at the edge of melt pool due to the recoil pressure

*2) Marangoni flow*: The thermocapillary force or Marangoni flows takes place whenever a non-uniform temperature exists along with an interface. In DED, the surface directly beneath the Laser beam is hotter than the surface near the edges of the melt pool. This temperature difference induces the gradients in surface tension causing the liquid metal to move from the lower surface tension region to the higher surface tension region. For 316L surface tension decreases with temperature, i.e., the surface tension at the cooler edge of the pool will be higher than that at the center. This creates a radially outward Marangoni flow that carries hot liquid to the edge of the pool, causing a shallow and wide melt pool.

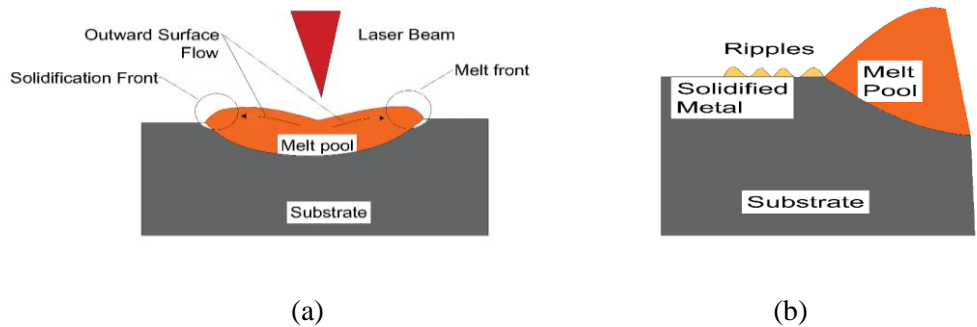

(a)          (b)

Fig. 12: (a) Outward Marangoni flow due to negative temperature gradient of surface tensions causing a shallow and wide melt pool in 316L, (b) Ripple formation at solidification front.

Marangoni flow can also be influenced by the amount of surface-active agents such as S and O present in the metal composition. A higher S content (> 150 ppm) can alter the flow patterns to create a deep and narrow melt pool.

### 5.3 Amplitude and spacing of ripple:

In a DED process for 316L, Marangoni (thermocapillary) flow patterns have the primary bearing on the melt pool flow patterns, especially near the solidification boundaries, and the recoil effect dominates the flow near the core of the melt pool. The observed ripple marks are largely informed by the Marangoni effect. According to Wie *et al's* [88] analytical model, the amplitude $a$ of the ripples may be expressed as

$$a = (1-K)\frac{\rho}{\gamma_m}\left[\nu\left(\left|\frac{d\gamma}{dT}\right|\frac{q_c h r_e}{k\mu}\right)^2\right]^{\frac{2}{3}} \quad\quad\quad \text{Eq.1}$$

where $K$ is the loss coefficient, $\rho$ is the density, $\gamma_m$ is the surface tension at melting temperature, $\nu$ is the kinematic viscosity, $\mu$ is the dynamic viscosity, $dT$ is the temperature difference between vertical walls, $q_c$ is the incident flux, $h$ is the fusion zone depth, $r_e$ is the rear length of the melt pool, $k$ is the liquid conductivity.

Wie et al. [88] also measured $a$ experimentally using Kosaka surfcoder SE 3000 and analytically mapped the average amplitude to a working parameter $p$ as

$$a = r \times p \qquad \text{Eq.2}$$

where $r$ is a proportionality constant determined for austenitic steels to be ~2.5, and it encapsulates the influence of the process parameters and certain material properties on the thermocapillary flow [88]. The working parameter $p$ is given by

$$p = \{[h_c(T_m - T_\infty) - \rho h_{sl} U]Qh^2\}^{\frac{2}{3}} \times 10^{-4} \qquad \text{Eq.3}$$

where, $h_c$ = 25 Wm$^{-2}$K$^{-1}$ is the heat transfer coefficient, $T_m - T_\infty = \Delta T_c$ = 1000 K, $\rho$ = 7040 kgm$^{-3}$ is the density, $h_{sl}$ = 250 Jkg$^{-1}$ is the latent heat for solidification, $h$ = 0.3×10$^{-3}$ m is the fusion zone depth and for our case $U \in$ (0.00084 – 0.0084) ms$^{-1}$ is the scan speed, $Q \in$ (300 – 600) W is the laser power.

After substituting the values in the working parameter equation, we get $p = 0.77 \times 10^{-4}$ and the amplitude of ripples of 5 μm. The value is consistent with the characteristic graph reported in the literature connecting the amplitude with the working parameter linearly [88]. It is evident from Eq. 3 that laser power and working parameter are directly proportional. As a result, the amplitude of the ripple increases as the laser power increases. Wie et al. [88] related the amplitude to the spacing (or the wavelength) $s$ of surface ripples as

$$a = cs \qquad \text{Eq.4}$$

where $a$ is the amplitude of ripples, and $c$ is a proportionality constant estimated empirically to be in the range of 0.015-0.03 for aluminum and steel alloys (e.g., Al 1100 and SS 304). The wavelength or spacing of the ripples calculated using Eq.4 range between $10^{-3}$ – $10^{-4}$ m, which is the same magnitude as that observed on the surface of aluminum and steel alloys [88].

Our experimental results capturing the combined effect of laser power and scan speed on the laser track width, ripple curvature, and ripple wavelength are summarized in Table 4. Here the ripple curvature was determined as mentioned in section 4. The wavelength of the ripple was estimated using Fourier analysis (FFT) of optical micrographs. Each micrograph was first converted to gray scale intensities value and then five laser tracks of length 2128 μm were selected and transformed into corresponding Fourier spectrum. The frequency with highest amplitude peak (as seen in Fig. 13 (a, b)) was chosen from five Fourier spectrum and the average ripple wavelength was calculated using the formula $\lambda = \frac{L}{f}$, where λ = ripple wavelength and L = length of the image along laser track and f = frequency of highest amplitude. The plots of Fig. 13 (a, b) show frequency portraits obtained at high and low energy densities. In high energy density case, a sharp low frequency component predominated whereas in low energy density, a diffused high frequency component predominated which explains the effect of other factors beyond marangoni convection influencing ripple geometry, which results in a small melt pool as the fluid in the melt pool becomes more stable. Furthermore, the plot of Fig. 13 (c, d) illustrates the variation of the amplitudes with the frequency estimated using gaussian process regression which showed the amplitude variation was less for low energy densities compared to high energy densities.

When the laser power is low (light color), and the scan speed is high (dark color) (samples 7 & 9) the ripple wavelengths tend to be low (light color) due to the relatively short interaction time between the laser and the deposited material on the surface. In contrast, when the laser intensity is high (dark color) and the scan speed is slow (light color) (sample 4), the interaction time and total energy transfer are high, leading to the longest ripple wavelength (dark color) among the samples observed. These experimental findings are consistent with the earlier findings [74], [75] noted in section 5.2 that an increase in the laser power and a decrease in the scan speed causes the ripple wavelength to increase.

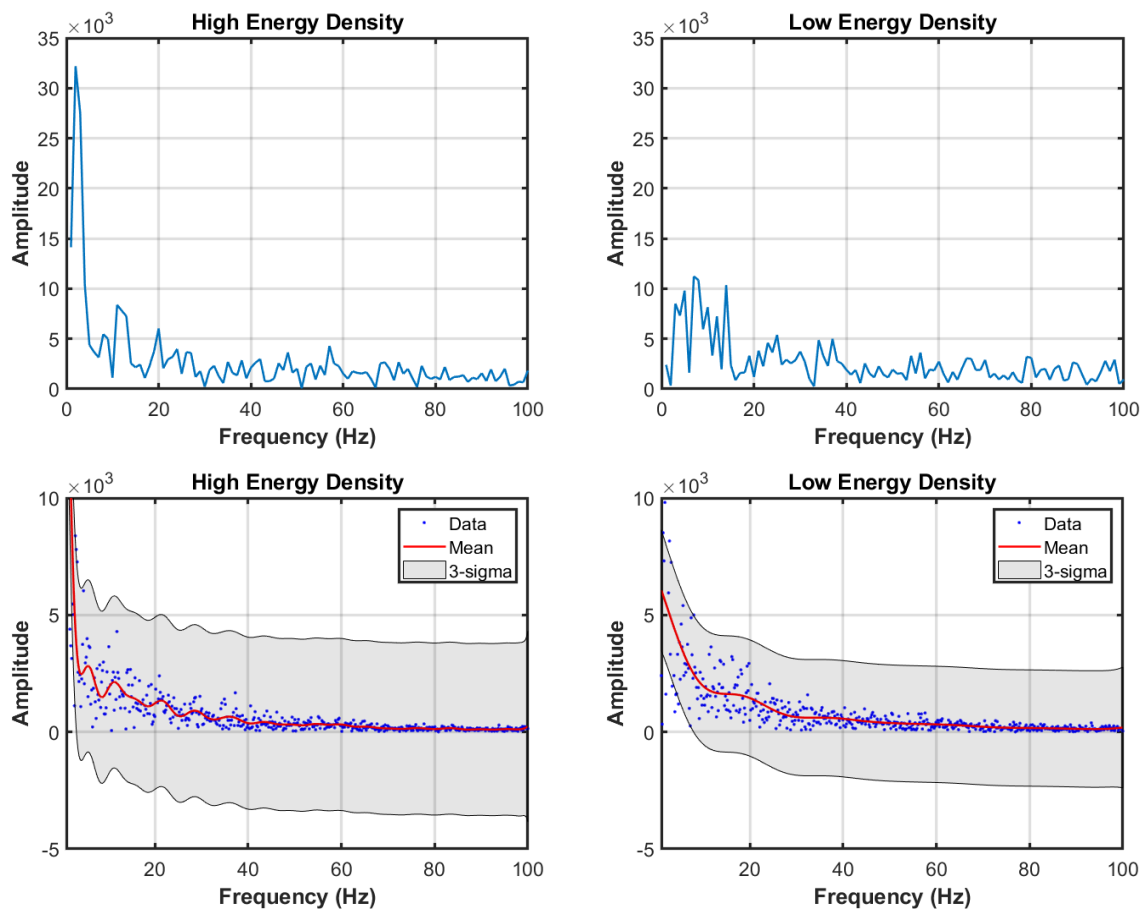

Fig.13: (a) Fourier spectra of high energy density sample, (b) Fourier spectra of low energy density sample, (c) amplitude variation of high energy density sample, (d) amplitude variation of low energy density sample

*Table 4* Observations of surface morphology under different process parameters

| PROCESS PARAMETER | | OBSERVATIONS | | |
|---|---|---|---|---|
| Laser power (W) | Scan speed (mm/sec) | Laser track width (µm) | Ripple curvature (µm) | Ripple wavelength (mm) |
| 562.5 | 8.46 | 468(95) | 514 (75) | 1.347 |
| 337 | 0.84 | 520(61) | 373(37) | 1.141 |
| 300 | 7.40 | 500(42) | 260(10) | 0.508 |
| 600 | 1.80 | 855(187) | 660(230) | 2.351 |
| 450 | 4.65 | 482(46) | 307(34) | 1.812 |
| 375 | 5.60 | 455(51) | 257(34) | 0.390 |

**Summary:**

The present study has investigated the surface microstructure of 316L printed in a DED process on an Optomec MTS 500 hybrid machine under different settings of the laser and scan speeds. The printed samples were polished to Ra < 30 nm, etched using Aqua Regia, and observed on an optical microscope. Notably, only five earlier works focused on the microstructure of the scan surface of DED-printed 316L samples. The following observations were made based on the micrographs taken at 4 different magnifications.

- The micrographs at 500 µm scale capture effects of the laser power and scan speed on the formation of ripples. While ripple formations have been studied in the context of welding processes, they have not received much attention in the DED literature. Based on the review of literature, the ripple formation in the DED process may be attributed primarily to the thermocapillary effect. From the experimental studies, it may be inferred that an increase of the laser power increases the ripple wavelength, the laser track width, and the ripple curvature. These observations are consistent with an earlier analytical model that predicts the spacing (wavelengths) of the ripples formed due to the thermocapillary effect.
- At 200 µm scale length the columnar/dendritic structures were evident at the laser track perimeter, and they were found to grow preferentially radially inward from melt pool boundary (ripple and laser track boundary).
- Evident from the micrographs at 100 and 50 µm scale lengths is that an increase in laser power transforms the dendritic structure from thin to coarse. These structures are determined by the spatial distribution of the temperature gradient G (°C/mm) and solidification growth rate R (mm/s) during the DED process.
- At low scan speed, cellular structure predominates and at higher scan speed the dendrite structure grains appear fine because the interaction time between the laser and powder becomes short and the solidification rate increases.

- The present investigation also lays a foundation for future work on the closed-loop control of a DED process, especially of the resulting microstructure. It is pertinent to note that any real-time control input essentially influences and should be synthesized based on the prediction or measurement of the morphology, especially the ripple and scan track formations and the resulting microstructure of the scan surface. In this context, the machine tools implementing the DED process are increasingly being instrumented with a combination of in-situ sensors, such as the melt pool cameras, temperature sensors, and various vibration and acoustic emission (AE) sensors [90]. A closed-loop control of the process can be achieved by controlling the ripple structure based on fusing the information contained in these various in-situ measurements.

- For several applications, the scan surface of a DED-printed part would be chosen as the load bearing and a contacting surface. Therefore, the microstructures and the morphological formations on the scan surface would influence the corrosion and wear resistance of the resulting component. The present study opens possibilities to perform a systematic analysis of behaviors and properties, such as corrosion resistance, hardness, wear resistance of the DED-printed component surfaces that bear the heterogeneous microstructure.

- Speeds and resolutions of conventional characterization techniques make them unwieldy to comprehensively map out the diverse microstructure patterns in DED. Prediction of such microstructures is possible by exploiting the signals obtained during machining studies post-printing to identify different sub-structures in microstructures [91]. Machining of the surface (e.g., via micro milling) allows for quick characterization of materials since the tool tips are of the same length scale as the microstructures found in 316L. By combining the information contained in images with sensor data acquired from the micromachining process, the framework may be expanded to quickly estimate phase boundaries in microstructures as well as process-to-material mapping.

**REFERENCES**

[1] T. Wohler, 'Additive manufacturing and 3D printing—state of the industry annual worldwide progress report 2014, Wohler's associates', *Inc Fort Collins CO*, 2013.
[2] T. Gornet, 'History of additive manufacturing non-SL systems Introduction of low-cost 3D printers', pp. 1–38, 2016.
[3] C. Atwood *et al.*, 'Laser Engineered Net Shaping (LENS(TM)): A Tool for Direct Fabrication of Metal Parts'.
[4] M. L. Griffith, L. D. Harwell, J. Romero, E. Schlienger, C. L. Atwood, and J. E. Smugeresky, 'Multi-material processing by LENS', *Solid Free. Fabr. Symp.*, pp. 387–393, 1997.
[5] H. Liu, Z. Hu, X. Qin, Y. Wang, J. Zhang, and S. Huang, 'Parameter optimization and experimental study of the sprocket repairing using laser cladding', *Int. J. Adv. Manuf. Technol.*, vol. 91, no. 9–12, pp. 3967–3975, 2017, doi: 10.1007/s00170-017-0066-y.
[6] Himanshu.Balhara, Abhishek.Kumar, S.Madhavan, 'Hard Facing of SS316LN by Direct Metal Deposition of Ni Based Wear Resistant Alloy', *Addit. Manuf. J.*, vol. 1, no. 3, pp. 38–42, 2019.
[7] R. Liu, Z. Wang, T. Sparks, F. Liou, and J. Newkirk, *13 - Aerospace applications of laser additive manufacturing*. Elsevier Ltd, 2017. doi: 10.1016/B978-0-08-100433-3.00013-0.

[8]  C. Selcuk, 'Laser metal deposition for powder metallurgy parts', *Powder Metall.*, vol. 54, no. 2, pp. 94–99, 2011, doi: 10.1179/174329011X12977874589924.

[9]  L. Costa and R. Vilar, 'Laser powder deposition', *Rapid Prototyp. J.*, vol. 15, no. 4, pp. 264–279, 2009, doi: 10.1108/13552540910979785.

[10] A. Yadollahi, D. Seely, B. Patton, N. Shamsaei, and S. M. Thompson, 'Mechanical and microstructural properties of LENS-produced AISI 316l stainless steel', *56th AIAAASCEAHSASC Struct. Struct. Dyn. Mater. Conf.*, no. January, pp. 1–8, 2015.

[11] M. A. Melia, H. A. Nguyen, M. Rodelas, and E. J. Schindelholz, 'Corrosion properties of 304L stainless steel made by directed energy deposition additive manufacturing', vol. 152, no. November 2018, pp. 20–30, 2019, doi: 10.1016/j.corsci.2019.02.029.

[12] M. Ma, Z. Wang, D. Wang, and X. Zeng, 'Optics & Laser Technology Control of shape and performance for direct laser fabrication of precision large-scale metal parts with 316L Stainless Steel', *Opt. Laser Technol.*, vol. 45, pp. 209–216, 2013, doi: 10.1016/j.optlastec.2012.07.002.

[13] B. Barkia *et al.*, 'Journal of Materials Science & Technology On the origin of the high tensile strength and ductility of additively manufactured 316L stainless steel : Multiscale investigation', vol. 41, pp. 209–218, 2020, doi: 10.1016/j.jmst.2019.09.017.

[14] M. Godec, S. Zaefferer, B. Podgornik, M. Šinko, and E. Tchernychova, 'Quantitative Multiscale Correlative Microstructure Analysis of Additive Manufacturing of Stainless Steel 316L Processed by Selective Laser Melting', *Mater. Charact.*, p. 110074, 2019, doi: 10.1016/j.matchar.2019.110074.

[15] J. D. Majumdar, A. Pinkerton, Z. Liu, I. Manna, and L. Li, 'Microstructure characterisation and process optimization of laser assisted rapid fabrication of 316L stainless steel', *Appl. Surf. Sci.*, vol. 247, pp. 320–327, 2005, doi: 10.1016/j.apsusc.2005.01.039.

[16] K. Zhang, S. Wang, W. Liu, and X. Shang, 'Characterization of stainless steel parts by Laser Metal Deposition Shaping', *Mater. Des.*, vol. 55, pp. 104–119, 2014, doi: 10.1016/j.matdes.2013.09.006.

[17] M. H. Farshidianfar, A. Khajepour, and A. P. Gerlich, 'Effect of real-time cooling rate on microstructure in Laser Additive Manufacturing', *J. Mater. Process. Technol.*, vol. 231, pp. 468–478, 2016, doi: 10.1016/j.jmatprotec.2016.01.017.

[18] F. Weng, S. Gao, J. Jiang, J. J. Wang, and P. Guo, 'A novel strategy to fabricate thin 316L stainless steel rods by continuous directed energy deposition in Z direction', *Addit. Manuf.*, vol. 27, no. March, pp. 474–481, 2019, doi: 10.1016/j.addma.2019.03.024.

[19] S. F. Yang, C. W. Li, A. Y. Chen, B. Gan, and J. F. Gu, 'Microstructure and corrosion resistance of stainless steel manufactured by laser melting deposition', *J. Manuf. Process.*, vol. 65, no. March, pp. 418–427, 2021, doi: 10.1016/j.jmapro.2021.03.051.

[20] Y. Yang, Y. Gong, C. Li, X. Wen, and J. Sun, 'Mechanical performance of 316 L stainless steel by hybrid directed energy deposition and thermal milling process', *J. Mater. Process. Technol.*, vol. 291, no. December 2020, p. 117023, 2021, doi: 10.1016/j.jmatprotec.2020.117023.

[21] D. Boisselier and S. Sankaré, 'Influence of Powder Characteristics in Laser Direct Metal Deposition of SS316L for Metallic Parts Manufacturing', *Phys. Procedia*, vol. 39, pp. 455–463, 2012, doi: 10.1016/j.phpro.2012.10.061.

[22] O. Freneaux, J. B. Poulet, O. Lepre, and G. Montavon, 'Coaxial nozzle for surface treatment by laser irradiation, with supply of materials in powder form'. Google Patents, 1995.

[23] K. Mahmood and A. J. Pinkerton, 'Direct laser deposition with different types of 316L steel particle : A comparative study of final part properties', *Eng. Manuf.*, vol. 227, no. 4, pp. 520–531, 2013, doi: 10.1177/0954405413475961.

[24] S. Wang, S. Zhang, C. H. Zhang, C. L. Wu, J. Chen, and M. B. Shahzad, 'Effect of $Cr_3C_2$ content on 316L stainless steel fabricated by laser melting deposition', *Vacuum*, vol. 147, pp. 92–98, 2018, doi: 10.1016/j.vacuum.2017.10.027.

[25] X. Wang, D. Deng, M. Qi, and H. Zhang, 'Influences of deposition strategies and oblique angle on properties of AISI316L stainless steel oblique thin-walled part by direct laser fabrication', *Opt. Laser Technol.*, vol. 80, pp. 138–144, 2016, doi: 10.1016/j.optlastec.2016.01.002.

[26] P. Guo, B. Zou, C. Huang, and H. Gao, 'Study on microstructure, mechanical properties and machinability of efficiently additive manufactured AISI 316L stainless steel by high-power direct laser deposition', *J. Mater. Process. Technol.*, vol. 240, pp. 12–22, 2017, doi: 10.1016/j.jmatprotec.2016.09.005.

[27] M. Mukherjee, 'Materialia Effect of build geometry and orientation on microstructure and properties of additively manufactured 316L stainless steel by laser metal deposition', *Materialia*, vol. 7, no. May, pp. 5–8, 2019, doi: 10.1016/j.mtla.2019.100359.

[28] F. Khodabakhshi, M. H. Farshidianfar, A. P. Gerlich, M. Nosko, V. Trembo, and A. Khajepour, 'E ff ects of laser additive manufacturing on microstructure and crystallographic texture of austenitic and martensitic stainless steels', *Addit. Manuf.*, vol. 31, no. June 2019, 2020, doi: 10.1016/j.addma.2019.100915.

[29] D. R. Feenstra, V. Cruz, X. Gao, A. Molotnikov, and N. Birbilis, 'E ff ect of build height on the properties of large format stainless steel 316L fabricated via directed energy deposition', *Addit. Manuf.*, vol. 34, no. March, p. 101205, 2020, doi: 10.1016/j.addma.2020.101205.

[30] Y. Balit, L. Joly, F. Szmytka, S. Durbecq, E. Charkaluk, and A. Constantinescu, 'Materials Science & Engineering A Self-heating behavior during cyclic loadings of 316L stainless steel specimens manufactured or repaired by Directed Energy Deposition', *Mater. Sci. Eng. A*, vol. 786, no. October 2019, p. 139476, 2020, doi: 10.1016/j.msea.2020.139476.

[31] Y. Balit, C. Guévenoux, A. Tanguy, M. V Upadhyay, E. Charkaluk, and A. Constantinescu, 'High resolution digital image correlation for microstructural strain analysis of a stainless steel repaired by Directed Energy Deposition', *Mater. Lett.*, vol. 270, p. 127632, 2020, doi: 10.1016/j.matlet.2020.127632.

[32] E. Tan Zhi'En, J. H. L. Pang, and J. Kaminski, 'Directed energy deposition build process control effects on microstructure and tensile failure behaviour', *J. Mater. Process. Technol.*, vol. 294, no. November 2020, p. 117139, 2021, doi: 10.1016/j.jmatprotec.2021.117139.

[33] A. Saboori *et al.*, 'Materials Science & Engineering A An investigation on the e ff ect of powder recycling on the microstructure and mechanical properties of AISI 316L produced by Directed Energy Deposition', *Mater. Sci. Eng. A*, vol. 766, no. August, p. 138360, 2019, doi: 10.1016/j.msea.2019.138360.

[34] M. Ma, Z. Wang, and X. Zeng, 'A comparison on metallurgical behaviors of 316L stainless steel by selective laser melting and laser cladding deposition', *Mater. Sci. Eng. A*, vol. 685, pp. 265–273, 2017, doi: 10.1016/j.msea.2016.12.112.

[35] J. Wang, 'Effect of External Magnetic Field on the Microstructure of 316L Stainless Steel Fabricated by Directed Energy Deposition', *Int. Mech. Eng. Congr. Expo.*, pp. 1–5, 2019.

[36] B. Zheng *et al.*, 'On the evolution of microstructure and defect control in 316L SS components fabricated via directed energy deposition', *Mater. Sci. Eng. A*, vol. 764, no. July, p. 138243, 2019, doi: 10.1016/j.msea.2019.138243.

[37] Y. Huang, M. Ansari, H. Asgari, and M. Hossein, 'Rapid prediction of real-time thermal characteristics , solidi fi cation parameters and microstructure in laser directed energy deposition ( powder- fed additive manufacturing )', *J. Mater. Process. Tech*, vol. 274, no. February, p. 116286, 2019, doi: 10.1016/j.jmatprotec.2019.116286.

[38] B. Rankouhi, K. M. Bertsch, G. M. De Bellefon, M. Thevamaran, D. J. Thoma, and K. Suresh, 'Materials Science & Engineering A Experimental validation and microstructure characterization of topology optimized , additively manufactured SS316L components', *Mater. Sci. Eng. A*, vol. 776, no. August 2019, p. 139050, 2020, doi: 10.1016/j.msea.2020.139050.

[39] J. Nie, L. Wei, D. Li, L. Zhao, Y. Jiang, and Q. Li, 'High-throughput characterization of microstructure and corrosion behavior of additively manufactured SS316L-SS431 graded material', *Addit. Manuf.*, vol. 35, no. February, p. 101295, 2020, doi: 10.1016/j.addma.2020.101295.

[40] X. Zhang, Y. Chen, and F. Liou, 'Fabrication of SS316L-IN625 functionally graded materials by powder-fed directed energy deposition', *Sci. Technol. Weld. Join.*, vol. 0, no. 0, pp. 1–13, 2019, doi: 10.1080/13621718.2019.1589086.

[41] S. M. Banait, C. P. Paul, A. N. Jinoop, H. Kumar, R. S. Pawade, and K. S. Bindra, 'Experimental investigation on laser directed energy deposition of functionally graded layers of Ni-Cr-B-Si and SS316L', *Opt. Laser Technol.*, vol. 121, no. March 2019, p. 105787, 2020, doi: 10.1016/j.optlastec.2019.105787.

[42] W. Woo *et al.*, 'Stacking Fault Energy Analyses of Additively Manufactured Stainless Steel 316L and CrCoNi Medium Entropy Alloy Using In Situ Neutron Diffraction', *Sci. Rep.*, pp. 2–4, 2020, doi: 10.1038/s41598-020-58273-3.

[43] P. Li, Y. Gong, Y. Xu, Y. Qi, Y. Sun, and H. Zhang, 'ScienceDirect Inconel-steel functionally bimetal materials by hybrid directed energy deposition and thermal milling : Microstructure and mechanical properties', *Arch. Civ. Mech. Eng.*, vol. 19, no. 3, pp. 820–831, 2019, doi: 10.1016/j.acme.2019.03.002.

[44] X. Zhang, T. Pan, Y. Chen, L. Li, Y. Zhang, and F. Liou, 'Additive manufacturing of copper-stainless steel hybrid components using laser-aided directed energy deposition', *J. Mater. Sci. Technol.*, vol. 80, pp. 100–116, 2021, doi: 10.1016/j.jmst.2020.11.048.

[45] J. C. Haley, J. M. Schoenung, and E. J. Lavernia, 'Modelling particle impact on the melt pool and wettability effects in laser directed energy deposition additive manufacturing', *Mater. Sci. Eng. A*, vol. 761, no. June, p. 138052, 2019, doi: 10.1016/j.msea.2019.138052.

[46] A. Aggarwal *et al.*, 'Role of impinging powder particles on melt pool hydrodynamics, thermal behaviour and microstructure in laser-assisted DED process: A particle-scale DEM – CFD – CA approach', *Int. J. Heat Mass Transf.*, vol. 158, p. 119989, 2020, doi: 10.1016/j.ijheatmasstransfer.2020.119989.

[47] N. Senin, A. Thompson, and R. K. Leach, 'Characterisation of the topography of metal additive surface features with different measurement technologies', *Meas. Sci. Technol.*, vol. 28, no. 9, 2017, doi: 10.1088/1361-6501/aa7ce2.

[48] N. Senin, A. Thompson, and R. Leach, 'Feature-based characterisation of signature topography in laser powder bed fusion of metals', *Meas. Sci. Technol.*, vol. 29, no. 4, 2018, doi: 10.1088/1361-6501/aa9e19.

[49] P. Tang *et al.*, 'The Formation of Humps and Ripples During Selective Laser Melting of 316l Stainless Steel', *JOM*, vol. 72, no. 3, pp. 1128–1137, Mar. 2020, doi: 10.1007/s11837-019-03987-7.

[50] T. Yang *et al.*, 'The influence of process parameters on vertical surface roughness of the AlSi10Mg parts fabricated by selective laser melting', *J. Mater. Process. Technol.*, vol. 266, pp. 26–36, Apr. 2019, doi: 10.1016/j.jmatprotec.2018.10.015.

[51] R. Hu *et al.*, 'Selective Laser Melting under Variable Ambient Pressure: A Mesoscopic Model and Transport Phenomena', *Engineering*, vol. 7, no. 8, pp. 1157–1164, Aug. 2021, doi: 10.1016/j.eng.2021.07.003.

[52] S. M. Banait, C. P. Paul, A. N. Jinoop, H. Kumar, R. S. Pawade, and K. S. Bindra, 'Experimental investigation on laser directed energy deposition of functionally graded layers of Ni-Cr-B-Si and SS316L', *Opt. Laser Technol.*, vol. 121, no. March 2019, p. 105787, 2020, doi: 10.1016/j.optlastec.2019.105787.

[53] K. Benarji, Y. R. Kumar, C. P. Paul, and A. N. Jinoop, 'Parametric investigation and characterization on SS316 built by laser-assisted directed energy deposition', *Mater. Des. Appl.*, vol. 234, no. 3, pp. 452–466, 2020, doi: 10.1177/1464420719894718.

[54] X. Wang, D. Deng, H. Yi, H. Xu, S. Yang, and H. Zhang, 'Influences of pulse laser parameters on properties of AISI316L stainless steel thin-walled part by laser material deposition', *Opt. Laser Technol.*, vol. 92, no. December 2016, pp. 5–14, 2017, doi: 10.1016/j.optlastec.2016.12.021.

[55] Z. Wang, T. A. Palmer, and A. M. Beese, 'Acta Materialia Effect of processing parameters on microstructure and tensile properties of austenitic stainless steel 304L made by directed energy deposition additive manufacturing', *Acta Mater.*, vol. 110, pp. 226–235, 2016, doi: 10.1016/j.actamat.2016.03.019.

[56] A. Saboori, F. Bosio, E. Librera, M. De Chirico, and S. Biamino, 'Accelerated Process Parameter Optimization For Directed Energy Deposition Accelerated Process Parameter Optimization for Directed Energy Deposition of 316L Stainless Steel', no. December 2019, 2018.

[57] G. Zhu, D. Li, A. Zhang, G. Pi, and Y. Tang, 'The influence of laser and powder defocusing characteristics on the surface quality in laser direct metal deposition', *Opt. Laser Technol.*, vol. 44, no. 2, pp. 349–356, 2012, doi: 10.1016/j.optlastec.2011.07.013.

[58] C. L. Wu, S. Zhang, C. H. Zhang, J. B. Zhang, Y. Liu, and J. Chen, 'E ff ects of SiC content on phase evolution and corrosion behavior of SiC- reinforced 316L stainless steel matrix composites by laser melting deposition', *Opt. Laser Technol.*, vol. 115, no. February, pp. 134–139, 2019, doi: 10.1016/j.optlastec.2019.02.029.

[59] N. Chen *et al.*, 'Microstructural characteristics and crack formation in additively manufactured bimetal material of 316L stainless steel and Inconel 625', *Addit. Manuf.*, vol. 32, no. February, p. 101037, 2020, doi: 10.1016/j.addma.2020.101037.

[60] N. Yang *et al.*, '3-D Direct Energy Deposition ( DED ) process induced material properties for the SS316L prototypes', 2016.
[61] Z. Sun, W. Guo, and L. Li, 'In-process measurement of melt pool cross-sectional geometry and grain orientation in a laser directed energy deposition additive manufacturing process', *Opt. Laser Technol.*, vol. 129, no. March, p. 106280, 2020, doi: 10.1016/j.optlastec.2020.106280.
[62] F. Khodabakhshi, M. H. Farshidianfar, A. P. Gerlich, M. Nosko, V. Trembo, and A. Khajepour, 'Materials Science & Engineering A Microstructure , strain-rate sensitivity , work hardening , and fracture behavior of laser additive manufactured austenitic and martensitic stainless steel structures', *Mater. Sci. Eng. A*, vol. 756, no. April, pp. 545–561, 2019, doi: 10.1016/j.msea.2019.04.065.
[63] K. Shah, A. Khan, S. Ali, M. Khan, and A. J. Pinkerton, 'Parametric study of development of Inconel-steel functionally graded materials by laser direct metal deposition', *Mater. Des.*, vol. 54, pp. 531–538, 2014, doi: 10.1016/j.matdes.2013.08.079.
[64] A. Yadollahi, N. Shamsaei, S. M. Thompson, and D. W. Seely, 'Effects of process time interval and heat treatment on the mechanical and microstructural properties of direct laser deposited 316L stainless steel', *Mater. Sci. Eng. A*, vol. 644, pp. 171–183, 2015, doi: 10.1016/j.msea.2015.07.056.
[65] H. Zhang *et al.*, 'Effect of Ni content on stainless steel fabricated by laser melting deposition', *Opt. Laser Technol.*, vol. 101, pp. 363–371, 2018, doi: 10.1016/j.optlastec.2017.11.032.
[66] F. Weng, S. Gao, J. Jiang, J. Wang, and P. Guo, 'A novel strategy to fabricate thin 316L stainless steel rods by continuous directed energy deposition in Z direction', *Addit. Manuf.*, vol. 27, no. November 2018, pp. 474–481, 2019, doi: 10.1016/j.addma.2019.03.024.
[67] W. Jin, W. Jin, M. Seob, J. Bae, and D. Sik, 'Repairing additive-manufactured 316L stainless steel using direct energy deposition', *Opt. Laser Technol.*, vol. 117, no. December 2018, pp. 6–17, 2019, doi: 10.1016/j.optlastec.2019.04.012.
[68] Y. Kok *et al.*, 'Anisotropy and heterogeneity of microstructure and mechanical properties in metal additive manufacturing : A critical review', *Mater. Des.*, vol. 139, pp. 565–586, 2018, doi: 10.1016/j.matdes.2017.11.021.
[69] T. Durejko, M. Pola, I. Kunce, P. Tomasz, K. J. Kurzyd, and Z. Bojar, 'Materials Science & Engineering A The microstructure , mechanical properties and corrosion resistance of 316 L stainless steel fabricated using laser engineered net shaping', *Mater. Sci. Eng. A*, vol. 677, pp. 1–10, 2016, doi: 10.1016/j.msea.2016.09.028.
[70] J. C. Betts, 'Journal of Materials Processing Technology The direct laser deposition of AISI316 stainless steel and Cr 3 C 2 powder', *J. Mater. Process. Technol.*, vol. 209, pp. 5229–5238, 2009, doi: 10.1016/j.jmatprotec.2009.03.010.
[71] D. K. Kim, W. Woo, E. Y. Kim, and S. H. Choi, 'Microstructure and mechanical characteristics of multi-layered materials composed of 316L stainless steel and ferritic steel produced by direct energy deposition', *J. Alloys Compd.*, vol. 774, pp. 896–907, 2019, doi: 10.1016/j.jallcom.2018.09.390.
[72] Y. Balit, E. Charkaluk, and A. Constantinescu, 'Digital image correlation for microstructural analysis of deformation pattern in additively manufactured 316L thin walls', *Addit. Manuf.*, vol. 31, no. May 2019, p. 100862, 2020, doi: 10.1016/j.addma.2019.100862.
[73] Y. Balit, E. Charkaluk, A. Constantinescu, and S. Durbecq, 'MICROSTRUCTURE AND PROCESS PARAMETERS FOR DIRECTED ENERGY DEPOSITION ADDITIVE', vol. 64, pp. 171–186, 2019.
[74] T. R. Anthony and H. E. Cline, 'Surface rippling induced by surface-tension gradients during laser surface melting and alloying', *J. Appl. Phys.*, vol. 48, no. 9, pp. 3888–3894, 1977, doi: 10.1063/1.324260.
[75] P. S. Wei, Y. H. Chen, J. S. Ku, and C. Y. Ho, 'Active Solute Effects on Surface Ripples in Electron-Beam Welding Solidification', *Metall. Mater. Trans. B*, vol. 34, no. August, 2003.
[76] A. S. Iquebal, S. Pandagare, and S. Bukkapatnam, 'Learning acoustic emission signatures from a nanoindentation-based lithography process: Towards rapid microstructure characterization', *Tribol. Int.*, vol. 143, no. October 2019, p. 106074, 2020, doi: 10.1016/j.triboint.2019.106074.
[77] M. Liu, A. Kumar, S. Bukkapatnam, and M. Kuttolamadom, 'A REVIEW OF THE ANOMALIES IN DIRECTED ENERGY DEPOSITION (DED) PROCESSES AND POTENTIAL SOLUTIONS'.

[78] C. Zhong, 'Experimental study of effects of main process parameters on porosity , track geometry , deposition rate , and powder efficiency for high deposition rate laser metal deposition', vol. 042003, 2015, doi: 10.2351/1.4923335.

[79] J. Hu, H. Guo, and H. L. Tsai, 'Weld pool dynamics and the formation of ripples in 3D gas metal arc welding', vol. 51, pp. 2537–2552, 2008, doi: 10.1016/j.ijheatmasstransfer.2007.07.042.

[80] J. Mazumder, J. Choi, K. Nagarathnam, J. Koch, and D. Hetzner, 'The direct metal deposition of H13 tool steel for 3-D components', *Jom*, vol. 49, no. 5, pp. 55–60, 1997, doi: 10.1007/BF02914687.

[81] D. S. Ertay, M. Vlasea, and K. Erkorkmaz, 'Thermomechanical and geometry model for directed energy deposition with 2D / 3D toolpaths', *Addit. Manuf.*, vol. 35, no. March, p. 101294, 2020, doi: 10.1016/j.addma.2020.101294.

[82] Z. Sun, W. Guo, and L. Li, 'Numerical modelling of heat transfer, mass transport and microstructure formation in a high deposition rate laser directed energy deposition process', *Addit. Manuf.*, vol. 33, no. March, p. 101175, 2020, doi: 10.1016/j.addma.2020.101175.

[83] J.D.Hunt, 'Steady State Columnar and Equiaxed Growth of Dendrites and Eutectic', *Mater. Sci. Eng.*, vol. 65, pp. 75–83, 1984.

[84] Y. Zhong, L. Liu, S. Wikman, D. Cui, and Z. Shen, 'Intragranular cellular segregation network structure strengthening 316L stainless steel prepared by selective laser melting', *J. Nucl. Mater.*, vol. 470, pp. 170–178, 2016, doi: 10.1016/j.jnucmat.2015.12.034.

[85] G. F. Vander Voort, G. M. Lucas, and E. P. Manilova, 'Metallography and Microstructures of Stainless Steels and Maraging Steels', *Metallogr. Microstruct.*, vol. 9, no. c, pp. 670–700, 2018, doi: 10.31399/asm.hb.v09.a0003767.

[86] N. Suutala and T. Takalo, 'Austenitic Stainless Steel Welds', vol. l, no. May, pp. 717–725, 1980.

[87] D. L. Cheever, 'Mechanism of Ripple Formation During Weld Solidification'.

[88] P. S. Wei, Chang C Y, and Chen C T, 'Surface Ripple in Electron-Beam Welding Solidification', *J. Heat Transf.*, vol. 118, no. November, 1996.

[89] P. S. Wei, 'Thermal Science of Weld Bead Defects : A Review', *J. Heat Transf.*, vol. 133, no. March 2011, 2011, doi: 10.1115/1.4002445.

[90] B. Botcha, A. S. Iquebal, and S. T. S. Bukkapatnam, 'Smart manufacturing multiplex', *Manuf. Lett.*, vol. 25, pp. 102–106, 2020, doi: 10.1016/j.mfglet.2020.08.004.

[91] A. S. Iquebal, B. Botcha, and S. Bukkapatnam, 'Towards rapid, in situ characterization for materials-on-demand manufacturing', *Manuf. Lett.*, vol. 23, pp. 29–33, 2020, doi: 10.1016/j.mfglet.2019.11.002.